\documentclass[journal]{IEEEtran}
\ifCLASSINFOpdf
  \usepackage[pdftex]{graphicx}
\else
  \usepackage[dvips]{graphicx}
\fi
\ifCLASSOPTIONcompsoc
  \usepackage[caption=false,font=normalsize,labelfont=sf,textfont=sf]{subfig}
\else
  \usepackage[caption=false,font=footnotesize]{subfig}
\fi
\usepackage{comment}
\usepackage{bm}
\usepackage{amsmath}
\usepackage{amsfonts}
\usepackage{xcolor}
\usepackage[normalem]{ulem}
\usepackage{physics}
\begin{document}

\newcommand{\Add}[1]{\textcolor{red}{#1}}
\DeclareRobustCommand{\Erase}{\bgroup\markoverwith{\textcolor{red}{\rule[.5ex]{2pt}{0.4pt}}}\ULon}
%\newcommand{\Erase}[1]{\if0{#1}\fi}


\title{Noncontact Haptic Rendering of Static Contact with Convex Surface Using Circular Movement of Ultrasound Focus on a Finger Pad}

\author{Tao~Morisaki,~\IEEEmembership{Member,~IEEE}
        Masahiro~Fujiwara,~\IEEEmembership{Member,~IEEE}
        Yasutoshi~Makino,
        and~Hiroyuki~Shinoda,~\IEEEmembership{Member,~IEEE}% <-this % stops a space
\thanks{Manuscript received xx; revised xx.}
\thanks{This work was supported in part by JSPS KAKENHI Grant Number 21J12305 and JST CREST JPMJCR18A2.}
\thanks{T. Morisaki is with the NTT Communication Science Laboratories, Nippon Telegraph and Telephone Corporation, Atsugi, Japan. E-mail: tao.morisaki@ntt.com}
\thanks{M. Fujiwara, Y. Makino, and H. Shinoda are with the Graduate School of Frontier Sciences, the University of Tokyo, Kashiwa-shi, Chiba, 277-8561, Japan (E-mail: fujimasa@nanzan-u.ac.jp; yasutoshi\_makino@k.u-tokyo.ac.jp; hiroyuki\_shinoda@k.u-tokyo.ac.jp).}% }% <-this % stops a space
}


\markboth{IEEE Transaction on Haptics}%
{Shell \MakeLowercase{\textit{et al.}}: Bare Demo of IEEEtran.cls for IEEE Journals}



\maketitle

\begin{abstract}
A noncontact tactile stimulus can be presented by focusing airborne ultrasound on the human skin. Focused ultrasound has recently been reported to produce not only vibration but also static pressure sensation on the palm by modulating the sound pressure distribution at a low frequency. This finding expands the potential for tactile rendering in ultrasound haptics as static pressure sensation is perceived with a high spatial resolution. In this study, we verified that focused ultrasound can render a static pressure sensation associated with contact with a small convex surface on a finger pad. This static contact rendering enables noncontact tactile reproduction of a fine uneven surface using ultrasound. In the experiments, four ultrasound foci were simultaneously and circularly rotated on a finger pad at 5~Hz. When the orbit radius was 3~mm, vibration and focal movements were barely perceptible, and the stimulus was perceived as static pressure. Moreover, under the condition, the pressure sensation rendered a contact with a small convex surface with a radius of 2~mm. The perceived intensity of the static contact sensation was equivalent to a physical contact force of 0.24~N on average, 10.9 times the radiation force physically applied to the skin. 
\end{abstract}


% \fi
\IEEEpeerreviewmaketitle

\newcommand{\rOne}{\mathrm{I}}
\newcommand{\rTwo}{\mathrm{I\hspace{-.1em}I}}
\section{Introduction}

\IEEEPARstart{A}{irborne} ultrasound tactile display (AUTD), which can present a noncontact tactile stimulus, is a promising tool for haptics since it does not require users to physically contact with any devices~\cite{rakkolainen2020survey}. An AUTD is a device with an array of independently controllable ultrasound transducers~\cite{hoshi2010noncontact,carter2013ultrahaptics}. AUTDs can focus ultrasound waves on arbitrary points in the air by controlling the phase of each transducer. At the focus, a nonnegative force called acoustic radiation force is generated~\cite{yosioka1955acoustic}, which conveys a noncontact tactile stimulus onto human skin. In addition to the investigation of the basis perception characteristics~\cite{wilson2014perception}, an AUTD has been used in various applications~\cite{rakkolainen2020survey}, such as human motion guidance~\cite{suzuki2019midair,yoshimoto2019midair,freeman2019haptiglow}, touchable midair image displays~\cite{monnai2014haptomime,romanus2019mid,morisaki2021midair}, remote communication system~\cite{makino2016haptoclone}, and automotive midair gesture interface~\cite{rumelin2017clicks, georgiou2017haptic, young2020designing, korres2020mid}, as the noncontact stimulus by AUTD does not obstruct a user’s movement and vision.

Recently, Morisaki et al. reported that AUTD can present not only vibratory sensations but also static pressure sensations~\cite{morisaki2021non}. A static pressure sensation is indispensable for tactile displays because the sensation is the main component of contact perception and is perceived with a higher resolution than vibratory sensations~\cite{johansson1983tactile}. However, in the conventional ultrasound haptics technique, a static pressure sensation is excluded from the presentable sensation of the AUTD. Ultrasound radiation force must be spatiotemporally modulated as it is less than several tens of mN~\cite{bolanowski1988four,hasegawa2018aerial,takahashi2019tactile,frier2018using}. This modulation has limited the tactile stimulus presented by the AUTD to a vibratory sensation. Morisaki et al. addressed this limitation and found that AUTD can present a static pressure sensation by repeatedly moving an ultrasound focus along the human skin at 5~Hz with a 0.2~mm spatial step width of the focus movement~\cite{morisaki2021non}. The focal trajectory was a 6~mm line, and the presentation location was a palm only.

In this study, we experimentally demonstrate that static pressure sensation by ultrasound can be evoked even at a finger pad. Moreover, we also show that by using a circular focal trajectory, the pressure sensation can render static contact with a small convex surface on the finger pad. The radius of the rendered convex surface is varied from 2 to 4~mm. Rendering static contact with such a small convex has been difficult for conventional ultrasound haptics because the perceptual resolution of vibratory sensations is lower than that of static pressure sensations~\cite{johansson1983tactile}. This contact sensation rendering enables the noncontact tactile reproduction of fine corrugated surfaces with a minimum spot size of several millimeters, which is equivalent to a spatial resolution of 1~cm. Using ultrasound, previous studies rendered an uneven surface (e.g., bumps and holes). However, in these studies, the contact sensation was not static as the finger and palm must be moved to perceive the rendered surface. Howard et al. and Somei et al. rendered an uneven surface by dynamically changing the intensity or position of the ultrasound focus according to hand movement~\cite{howard2019investigating,somei2022spatial}.

In the experiment, a focus rotating in a circle at 5 Hz is presented to a finger pad, and the radius of the trajectory is varied from 2 to 6~mm. We evaluate the intensity of the vibratory and movement sensations of the focus produced by the presented stimulus. We also evaluated the curvature of the tactile shape (i.e., flat, convex, or concave) perceived on the finger pad. Moreover, we examine the optimal ultrasound focus shape for creating a perfect static pressure sensation.

\section{Related Works}
In this section, we summarize previous studies on point stimulation and haptic shape rendering using ultrasound.

\subsection{Vibratory and Static Pressure Sensation by Ultrasound}
Two methods have been employed to create a single point vibrotactile sensation: Amplitude Modulation (AM)~\cite{hasegawa2018aerial} and Lateral Modulation (LM)~\cite{takahashi2018lateral,takahashi2019tactile}. AM is a stimulation method wherein the amplitude of the presented radiation pressure is temporally modulated~\cite{hasegawa2018aerial}. In LM, a vibratory stimulus is presented by periodically moving a single stimulus point (ultrasound focus) along the skin surface with constant pressure~\cite{takahashi2018lateral,takahashi2019tactile}. Takahashi et al. showed that the perceptual threshold of the LM was lower than that of the AM ~\cite{takahashi2018lateral,takahashi2019tactile}. The employed focal trajectory was a line and circle with representative lengths of a few millimeters. Spatiotemporal Modulation (STM) has also been used to create a larger trajectory of a moving focus~\cite{frier2018using,frier2019sampling}. Frier et al. presented a circular STM with circumferences of 4--10~cm~\cite{frier2018using}.

A static pressure sensation can be produced by a low-frequency LM stimulus with a fine spatial step width of the focal movement. Morisaki et al. presented a static pressure sensation using an LM at 5~Hz with a step width of 0.2~mm~\cite{morisaki2021non}. The focal trajectory was a 6~mm line. Under this condition, the vibratory sensation included in the LM was suppressed to 5\% in a subjective measure, and the perceived intensity was comparable to 0.21~N physical pushing force on average. A similar phenomenon, evoking presser sensation by vibration, has been also confirmed with a vibrator~\cite{konyo2005tactile}. The pressure sensation by ultrasound has been presented only on the palm, and whether the pressure sensation can be evoked on a finger pad has not been confirmed. This study aims to present the pressure sensation to a finger pad. Morisaki et al. and Somei et al. presented a low frequency-fine step LM stimulus to a finger pad. However, they did not evaluate its tactile feeling~\cite{morisaki2021midair,somei2022spatial}.

\subsection{Rendering Haptic Shape Using Ultrasound}
Several studies have presented symbolic two-dimensional haptic shapes, such as a line and circle on the palm using AUTDs. To render them, Korres and Eid and Ruttern et al.~\cite{korres2016haptogram, rutten2019invisible} used AM with multiple foci. Marti et al. used STM, wherein the focal trajectory is the perimeter of the target shape~\cite{marti2021mid}. Hajas et al. also drew such 2D tactile shapes by periodically moving an amplitude-modulated focus on the perimeter~\cite{hajas2020mid}. Mulot et al. drew a curved line to the palm using STM and evaluated whether its curvature can be discriminated~\cite{mulot2021dolphin,mulot2021curvature}.

Moreover, AUTD has been used for tactile reproduction of contact between 3D objects and hands. Inoue et al. presented a 3D static haptic image using an ultrasound standing wave~\cite{inoue2015active}. Long et al. rendered the contact shape with a virtual 3D object to a palm using multiple ultrasound foci~\cite{long2014rendering}. Matsubayashi calculated the contact area between a finger and a virtual 3D object and rendered this area to a finger pad by presenting an LM whose focal trajectory was the perimeter of the calculated contact area~\cite{matsubayashi2019direct,matsubayashi2019display}. These studies aimed to reproduce the macroscopic shape of a 3D object and did not reproduce contact shape with a fingertip-sized small convex surface, as in this study. Moreover, static pressure sensations were not presented in these studies. Long et al. used AM at 200 Hz~\cite{long2014rendering} and Matsubayashi et al. LM at 100 Hz~\cite{matsubayashi2019direct,matsubayashi2019display}. The static haptic image presented by Inoue et al. was not modulated, but the participants had to keep moving their hands to perceive its tactile sensations~\cite{inoue2015active}.

Several studies have reproduced uneven surfaces using AUTD. Howard et al. presented three tactile shapes to a palm: bump, hole, and flat, by dynamically changing the stimulus intensity for the hand position~\cite{howard2019investigating}. Somei et al. presented a convex surface sensation to a finger pad by changing the stimulus position for the finger position~\cite{somei2022spatial}. Perceived tactile shapes using these methods require active finger or hand movement. In contrast, this study render a static convex shape to a stationary finger.

%Fig.~\ref{fig:Fig/Principle/AUTD.pdf} shows the AUTD. 

\section{Stimulus Design\label{sec: Stimulus Design}}
\newcommand{\upperTrans}{_{\mathrm{trans}}}
\newcommand{\upperFocus}{^{\mathrm{F}}}
\newcommand{\Center}{\bm{c}}
\newcommand{\Point}{\bm{r}}
\newcommand{\UnitX}{\Point_\mathrm{a}}
\newcommand{\UnitY}{\Point_\mathrm{b}}
\newcommand{\UnitZ}{\Point_\mathrm{c}}
\newcommand{\FocusPosition}{\Point}
\newcommand{\SoundPressure}{P}
\newcommand{\RadiationForce}{F}
\newcommand{\TransNumTotal}{N\upperTrans}
\newcommand{\myTime}{t}
\newcommand{\FocusNum}{i}
\newcommand{\FocusNumTotal}{N_{\mathrm{focus}}}

\newcommand{\StimulusRadius}{A}
\newcommand{\CenterPosition}{\FocusPosition_{\mathrm{cnt}}}
\newcommand{\CircleX}{x_\mathrm{c}}
\newcommand{\CircleY}{y_\mathrm{c}}
\newcommand{\CircleZ}{z_\mathrm{c}}
\newcommand{\SampleNumTotal}{N}
\newcommand{\SampleNum}{j}
\newcommand{\LMAngle}{\theta_\SampleNum}
\newcommand{\CirclePositionEach}{\FocusPosition_{\mathrm{c},\SampleNum}}
\newcommand{\LmFreq}{f^\mathrm{LM}}
\newcommand{\LmStep}{d^\mathrm{LM}}
\newcommand{\MultiSpace}{d}
\newcommand{\MultiSpacePoint}{l}
\newcommand{\myZ}{z}
\newcommand{\DwellTime}{\myTime_\mathrm{d}}

\newcommand{\phase}{\bm{\phi}}
\newcommand{\phasEach}{\bm{\phi}_\FocusNum}

\begin{figure}[!t]
    \centering
    \includegraphics[width=0.9\columnwidth]{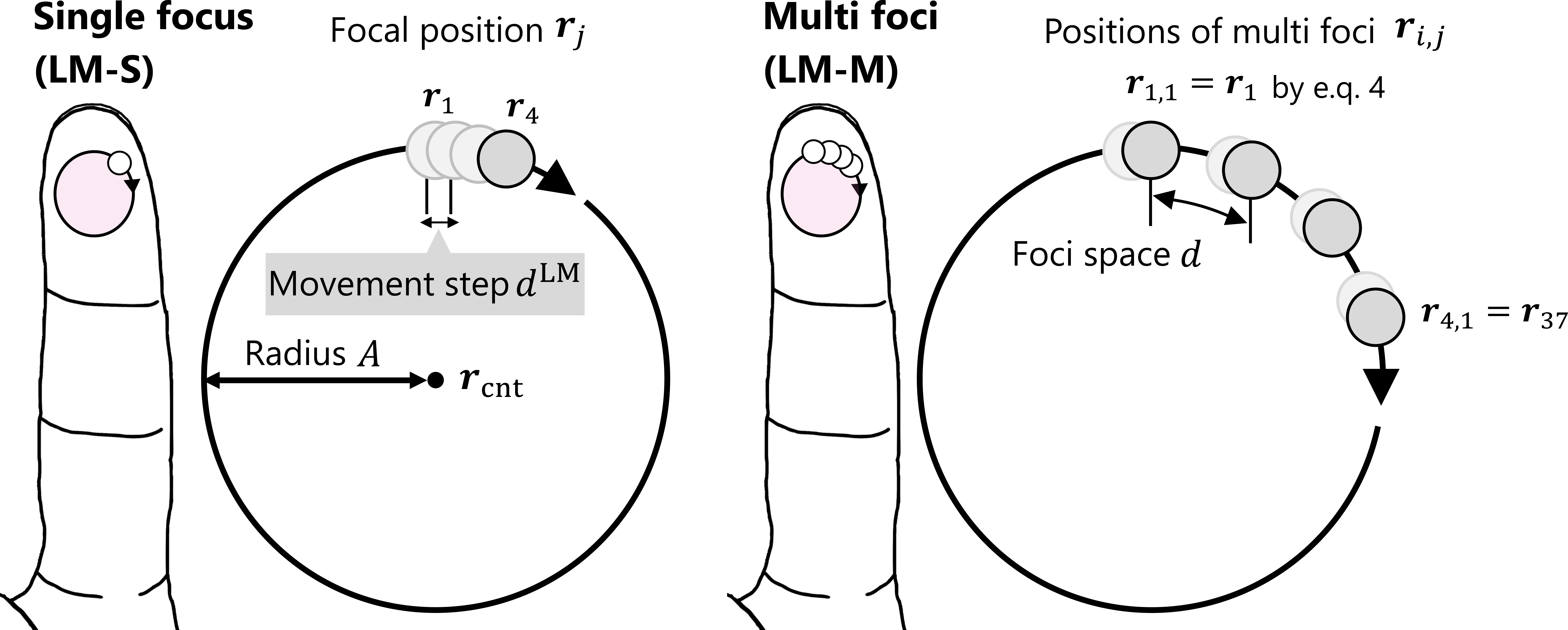}
    \caption{Schematic of LM-S (single focus) stimulus and LM-M (multi foci) stimulus. In the LM-S, a single focus is periodically moved in a circle on a finger pad. In the LM-M, multiple foci, placed along with the circular trajectory, are simultaneously rotated. The LM-M schematic also shows the example  $\FocusPosition_{\FocusNum,\SampleNum}$--$\FocusPosition_{\SampleNum}$ conversion with the foci number $\FocusNumTotal = 4$, the foci space $\MultiSpace = 3$ mm, and the step width of focus movement $\LmStep = 0.22$ mm.}
    \label{fig:Fig/Principle/LM_Stimulus.pdf}
\end{figure}

 \begin{figure}[t]
    \centering
    \includegraphics[width=0.9\columnwidth]{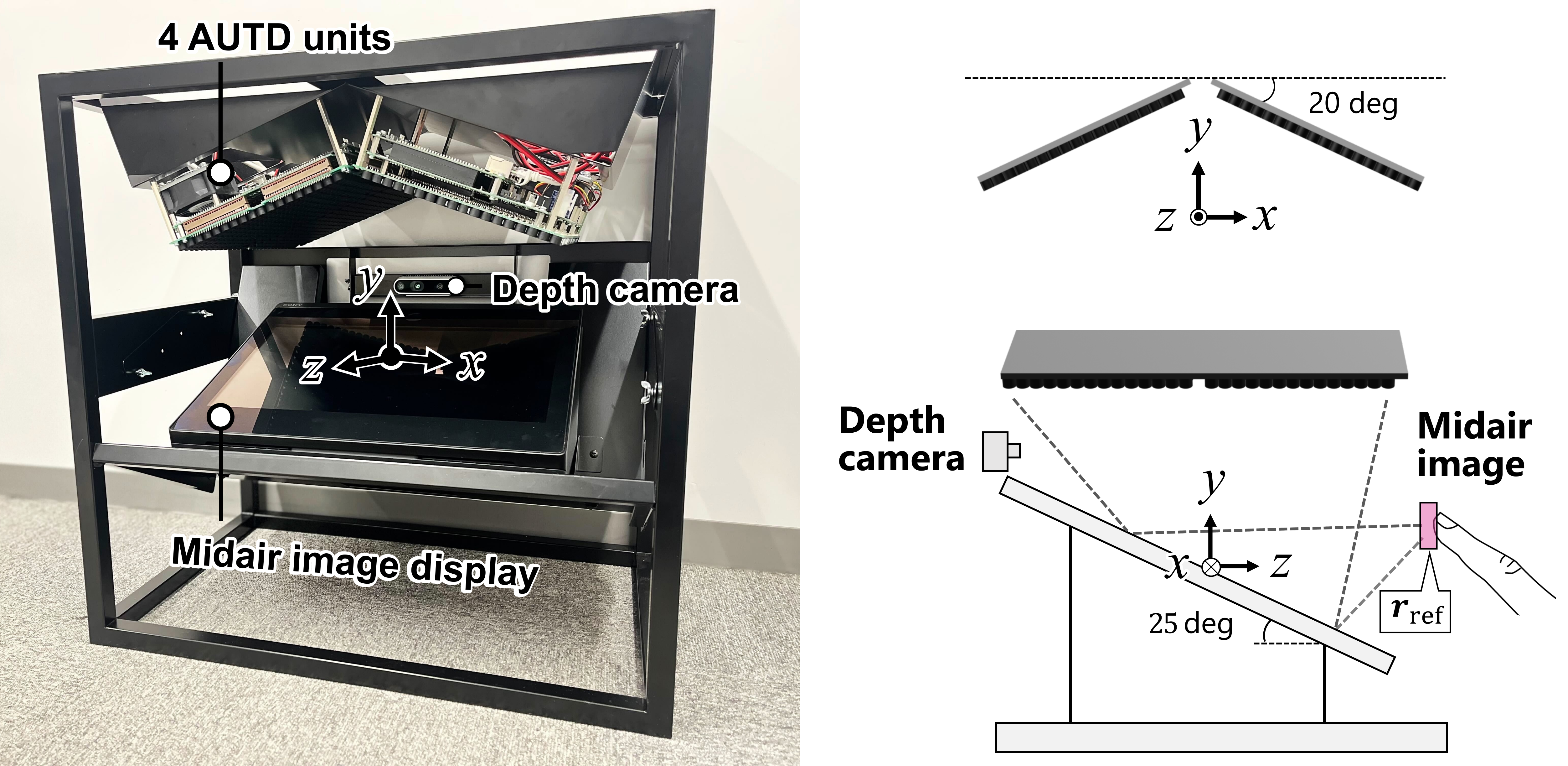}
    \caption{Experimental equipment used in all subject experiments in this study. This equipment presents a midair image marker. An ultrasound tactile stimulus (LM stimulus) is presented when the finger of a participant touches this marker. The marker was used to indicate a finger position to a participant.}
    \label{fig:Fig/Equipment/ExperimentalSetupActual.pdf}
\end{figure}   

\subsection{Overview}
In this section, we propose and describe two stimulus methods: LM-single focus (LM-S) and LM-multi foci (LM-M). In the subject experiment, we compared and evaluated them to investigate whether they could render a static contact sensation with a convex surface. Fig.~\ref{fig:Fig/Principle/LM_Stimulus.pdf} shows a schematic of these stimulus methods. In LM-S, a single ultrasound focus is periodically moved in a circle on the finger pad. The LM-S has been used in previous studies~\cite{frier2018using, takahashi2019tactile, matsubayashi2019direct}; however, these studies have not evaluated whether this stimulus can produce static pressure and static contact sensations. In the LM-M, multiple ultrasound foci were simultaneously presented and periodically moved in a circle. The foci were placed along the circular focal trajectory so that they were in close proximity. The distance between foci $\MultiSpace$ was fixed at 3 mm in the experiments. Previous study has proposed STM stimulus using multi foci simultaneously, similar to the LM-M~\cite{plasencia2020gs, shen2023multi}; however, it was not used to render a haptic shape. Plasencia et al. only performed physical measurements~\cite{plasencia2020gs}. Shen et al. evaluated the emotional effect caused by the multi-point STM~\cite{shen2023multi}. Moreover, the STM scale greatly differs from the LM-M. The circumference of the multi-point STM was over 24 cm, but that of the LM-M was under 3.8 cm.

In the experiment, the amplitude of each transducer was set to maximum and the driving phase for presenting the LM-M stimulus was calculated using a linear synthesis scheme. Let $\phasEach \in \mathbb{R}^{\TransNumTotal}$ be the phase for presenting each focus in the LM-M, and the phase for simultaneously presenting multiple foci $\phase \in \mathbb{R}^{\TransNumTotal}$ is expressed as follows:
\begin{IEEEeqnarray}{rCl}
\phase = \sum_{\FocusNum}^{\FocusNumTotal}\phasEach,
\end{IEEEeqnarray}
where $\FocusNum \in \{1,...\FocusNumTotal\}$ is the index number of multiple foci, $\FocusNumTotal$ is the total number of multiple foci, and $\TransNumTotal$ is the total number of transducers.

\subsection{Formulation}
First, we formulated a focus movement for the LM-S stimulus. The focus position in LM-S $\FocusPosition_{\SampleNum} \in \mathbb{R}^3$ is given by the following:
\begin{IEEEeqnarray}{rCl}
\FocusPosition_{\SampleNum} &=& \CenterPosition + \StimulusRadius(\cos{\LMAngle}\UnitX+\sin{\LMAngle}\UnitY) + \myZ_\SampleNum\UnitZ,\label{eq:SinglePointLM}\\
\LMAngle &=& \frac{2\pi}{N}(\SampleNum-1),
\end{IEEEeqnarray}
where $\SampleNum \in \{1,...\SampleNumTotal\}$ is the index of the focus position, $\SampleNumTotal$ is the total number of focus positions in one cycle of the LM, $\CenterPosition \in \mathbb{R}^3$ is the center of the focal trajectory, and $\StimulusRadius$ is the radius of the trajectory. $\UnitX$, $\UnitY$, and $\UnitZ$ are unit vectors whose origin is at $\CenterPosition$ and parallel to the x-, y-, and z-axis, respectively. The value of $\myZ_\SampleNum$ was determined using the measured finger depth position. Based on these definitions, the step width of the focus movement is $\LmStep = \frac{2\pi\StimulusRadius}{\SampleNumTotal}$. The index of focus position $\SampleNum$ changes after the dwell time of focus $\DwellTime$. Dwell time was $\DwellTime = \frac{1}{\SampleNumTotal\LmFreq}$ if the frequency of the LM stimulus is $\LmFreq$. 

Second, we formulated the LM-M stimulus. Let $\FocusPosition_{\FocusNum,\SampleNum} \in \mathbb{R}^3$ be the focus position on the LM trajectory of the $\FocusNum$-th focus among the foci presented simultaneously. $\FocusPosition_{\FocusNum,\SampleNum}$ is chosen from $\FocusPosition_{\SampleNum}$, which is the position discretized with $\LmStep$, such that the motion step width of the multi foci is fixed to $\LmStep$. The conversion from $\FocusPosition_{\SampleNum}$ to $\FocusPosition_{\FocusNum,\SampleNum}$ is expressed as follows:
\begin{IEEEeqnarray}{rCl}
\FocusPosition_{\FocusNum,\SampleNum} &=& \FocusPosition_{\SampleNum + (\FocusNum -1)\MultiSpacePoint}, \label{eq:MultiPointLM}\\
\MultiSpacePoint &=& \left\lfloor\frac{\MultiSpace}{\LmStep}\right\rfloor,
\end{IEEEeqnarray}
where $\MultiSpace$ is the distance between the multi foci and $\MultiSpacePoint$ is the index number calculated from the $\MultiSpace$. $\MultiSpacePoint$ is an integer, and the decimal point is rounded down. 

%d/d^LMは14だった
Here we show the example of the conversion with the foci number $\FocusNumTotal = 4$, the foci space $\MultiSpace = 3$ mm, the step width of focus movement $\LmStep = 0.22$ mm. When $\myTime = 0$, the four foci positions in LM-M ($\FocusPosition_{\FocusNum,\SampleNum}$) are $\FocusPosition_{1, 1}$, $\FocusPosition_{2, 1}$, $\FocusPosition_{3, 1}$, and $\FocusPosition_{4, 1}$. These positions are converted to $\FocusPosition_{1}$, $\FocusPosition_{15}$, $\FocusPosition_{29}$, and $\FocusPosition_{37}$, respectively by eq.~\ref{eq:MultiPointLM}.

\section{Experimental Equipment\label{sec: Experimental Equipment}}
\newcommand{\PointRot}{\bm{r}_\mathrm{ref}}
\newcommand{\PointBeforeRot}{\bm{r}_\mathrm{org}}
\newcommand{\DisplayTheta}{\theta}
\newcommand{\PointReference}{\bm{r}_\mathrm{p}}
\newcommand{\NormalVector}{\bm{n}}
\newcommand{\Inclination}{m}
\newcommand{\MirrorMatrix}{M}

\begin{figure}[t]
    \centering
    \includegraphics[width=1\columnwidth]{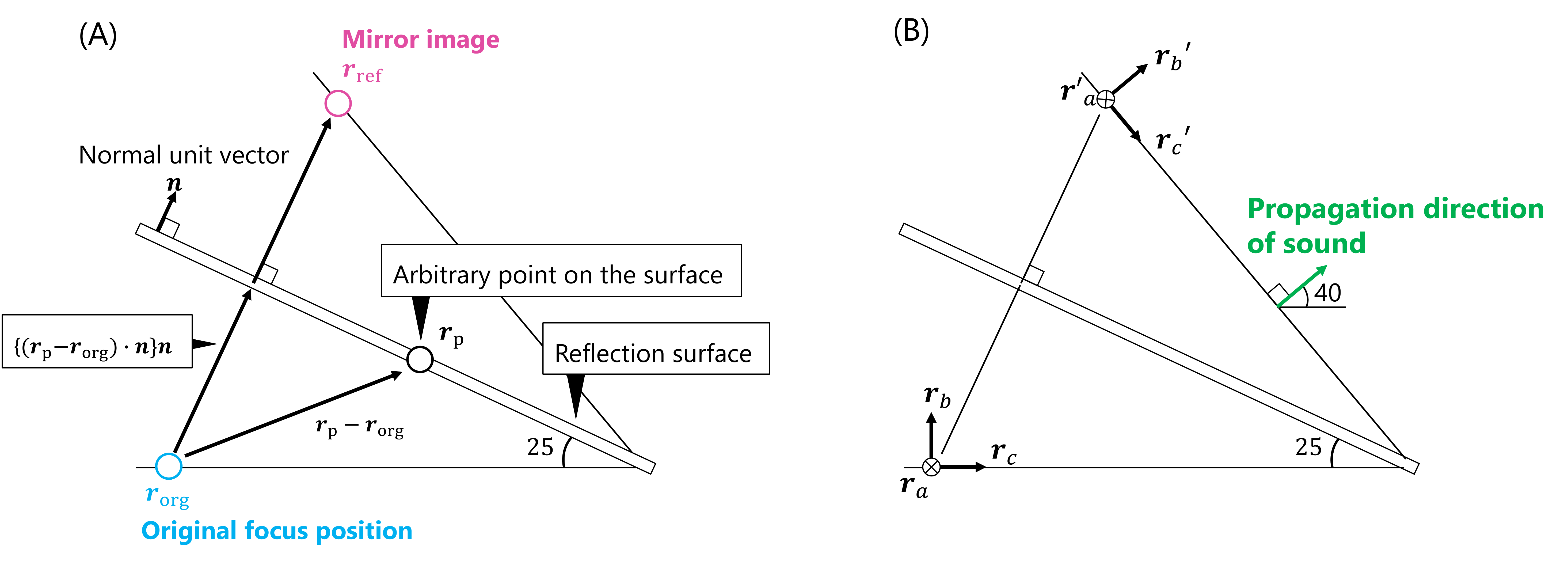}
    \caption{A) Schematic of mirror image conversion of focus position. When presenting ultrasound focus at $\PointBeforeRot$, the focus is formed at $\PointRot$ as the ultrasound reflects on the display surface. B) Example of unit vector ($\UnitX, \UnitY, \UnitZ$) and converted unit vectors ($\UnitX', \UnitY', \UnitZ'$) used to define the LM trajectory before and after conversion, respectively. The propagation direction of reflected ultrasound is parallel to the $\UnitY'$.}
    \label{fig:Fig/Equipment/SchematicReflection.pdf}
\end{figure}

In this section, we describe the experimental equipment that presents a midair image with noncontact tactile feedback. This equipment was used in all the subject experiments conducted in this study. 

\subsection{System Overview}
Fig.~\ref{fig:Fig/Equipment/ExperimentalSetupActual.pdf} shows the experimental equipment and its coordinate system. This system consists of the four AUTDs, a midair image display (ELF-SR1 Spatial Reality Display, SONY), and a depth camera (RealSense D435, Intel) used to measure the finger position. The one AUTD unit was equipped with 249 ultrasound transducers operating at 40 kHz (TA4010A1, NIPPON CERAMIC Co., Ltd.)~\cite{suzuki2021autd3}. Each AUTD communicated via the EtherCAT protocol and was synchronously driven. The AUTDs were tilted 20 deg around the z-axis to concentrate the acoustic energy to the center of the system (stimulus position). In the experiments, we used the midair image display to instruct participants where to put their fingers. The coordinate system is a right-handed system whose origin is the center of the surface of the image display.

Throughout all the experiments, the system presented a 1 $\times$ 1~cm image marker at (0, 30, 30)~mm. Ultrasound waves were output from the AUTDs when participants placed their fingertips on the marker. The presented ultrasound wave reflected on the surface of the image display and then focused on the finger pad. The schematic of the reflection is shown in Fig.~\ref{fig:Fig/Equipment/SchematicReflection.pdf}. The position of the reflected ultrasound focus $\PointRot \in \mathbb{R}^3$ can be calculated as the mirror image of the original focus position $\PointBeforeRot \in \mathbb{R}^3$ which is expressed as follows:
\begin{IEEEeqnarray}{rCl}
\PointRot &=& \PointBeforeRot+2((\PointReference-\PointBeforeRot)\cdot\NormalVector)\NormalVector\label{eq:PointRot},
\end{IEEEeqnarray}
where $\NormalVector$ is the normal vector of the display surface (reflective surface), and $\PointReference$ is an arbitrary point on the display surface. From eq.~\ref{eq:SinglePointLM} and \ref{eq:PointRot}, the focal trajectory in LM stimulus after the reflection is described as follows: 
\begin{IEEEeqnarray}{rCl}
\FocusPosition_{\SampleNum} &=& \PointRot + \StimulusRadius(\cos{\LMAngle}\UnitX'+\sin{\LMAngle}\UnitY') + \myZ_\SampleNum\UnitZ',\label{eq:reflectedLM}
\end{IEEEeqnarray}
where $\UnitX'$ is parallel shifted $\UnitX$. $\UnitY'$ and $\UnitZ'$ are $\UnitY$ and $\UnitZ$ tilted -50 deg around $\UnitX'$, respectively.

Fig.~\ref{fig:Fig/Equipment/SimulatedRadiationPressure_simple.pdf} shows the simulated radiation force distribution of a single focus and multi foci. The single focus position $(0, 0, 220)$ mm is shown in the transducer setup (Fig.~\ref{fig:Fig/Equipment/SetupAUTD.pdf}) as a cross mark. The amplitude of each transducer was set to 1 a.u. Sound wave reflection was not considered. The multi foci were placed according to the configuration of LM-M with $\StimulusRadius = 6$ mm. The LM center $\CenterPosition$ was also $(0, 0, 220)$ mm. The large and small white circle drawn in the LM-M simulation indicates the focal trajectory of LM-M and each focus position, respectively. The transducers were modeled as point sound sources and the directivity was determined by the spec sheets of the actually used transducers (described in Section~\ref{sec: Experimental Equipment}).

\subsection{Algorithm for Presenting LM Stimulus}
\begin{figure}[t]
    \centering
    \includegraphics[width=0.8\columnwidth]{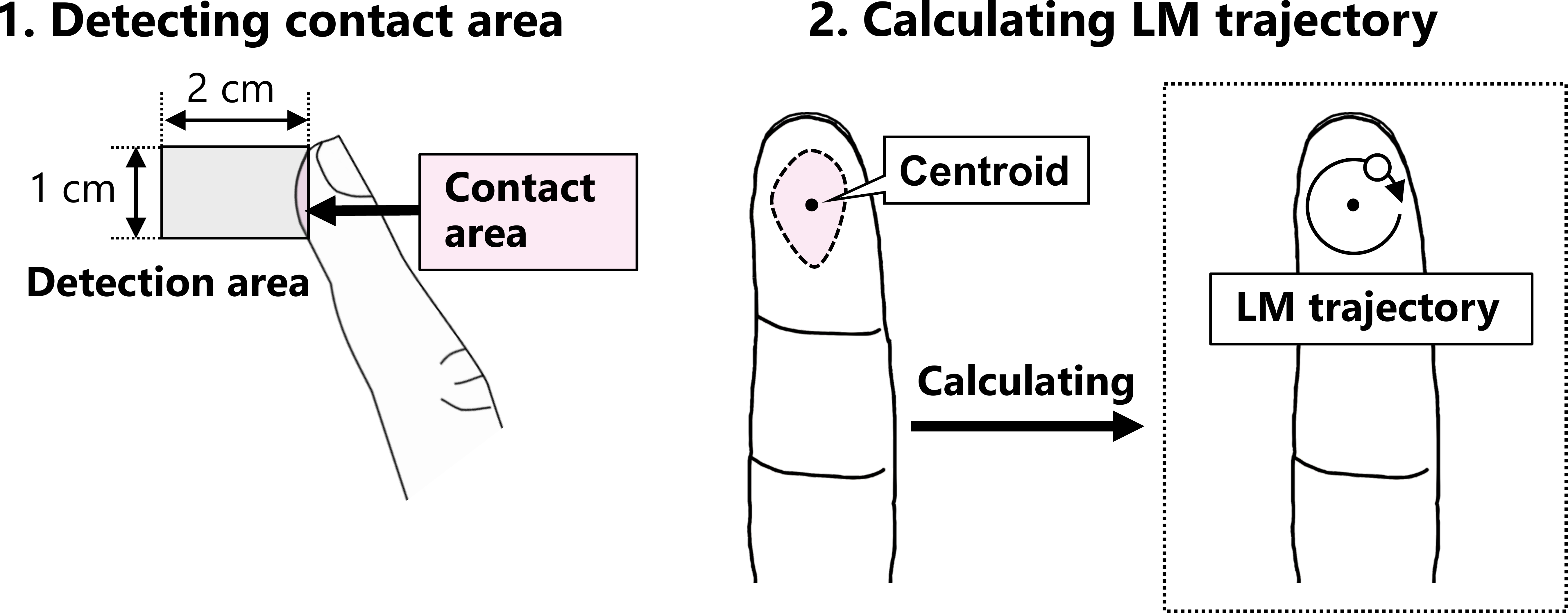}
    \caption{Algorithm for presenting LM stimulus. The size of the detection area is $1\times1\times2$~cm. The part of the finger within this detection area is measured as the contact area, and the focal trajectory of the LM stimulus is calculated using this area. The center of the LM is the centroid of the contact area.}
    \label{fig:Fig/Equipment/DetectionFlow.pdf}
\end{figure}
In the system, there are three processes for presenting a circular LM stimulus to the finger pad of the participant. Fig.~\ref{fig:Fig/Equipment/DetectionFlow.pdf} illustrates the presentation process. First, the system detects the contact area between a participant's finger and midair image marker using a depth camera. The size of the image marker is $1\times1\times0.5$~cm. However, to measure the contact position stably, we used the area from the surface of the image marker to 2~cm behind ($1\times1\times2$~cm) for the contact detection. Part of the finger within the detection area was measured as the contact area. Second, the system calculated the focal trajectory for the circular LM stimulus using eq.~\ref{eq:SinglePointLM} or eq.~\ref{eq:MultiPointLM}. The center position of the LM stimulus $\CenterPosition$ was the centroid of the detected contact area. The measured depth map of the fingertip surface was used for the z-position of the focal trajectory. Third, the focus is presented and moved along with the calculated trajectory at a pre-specified frequency. In this algorithm, the $\CenterPosition$ is asynchronously updated with the focus position at 90 fps. A Gaussian filter was applied to the calculated $\CenterPosition$ of 10 frames to suppress the measurement error of the depth camera.

\subsection{Measurement of Radiation Force\label{sec: Measurement of Radiation Force}}
\begin{figure}[t]
    \centering
    \includegraphics[width=0.7\columnwidth]{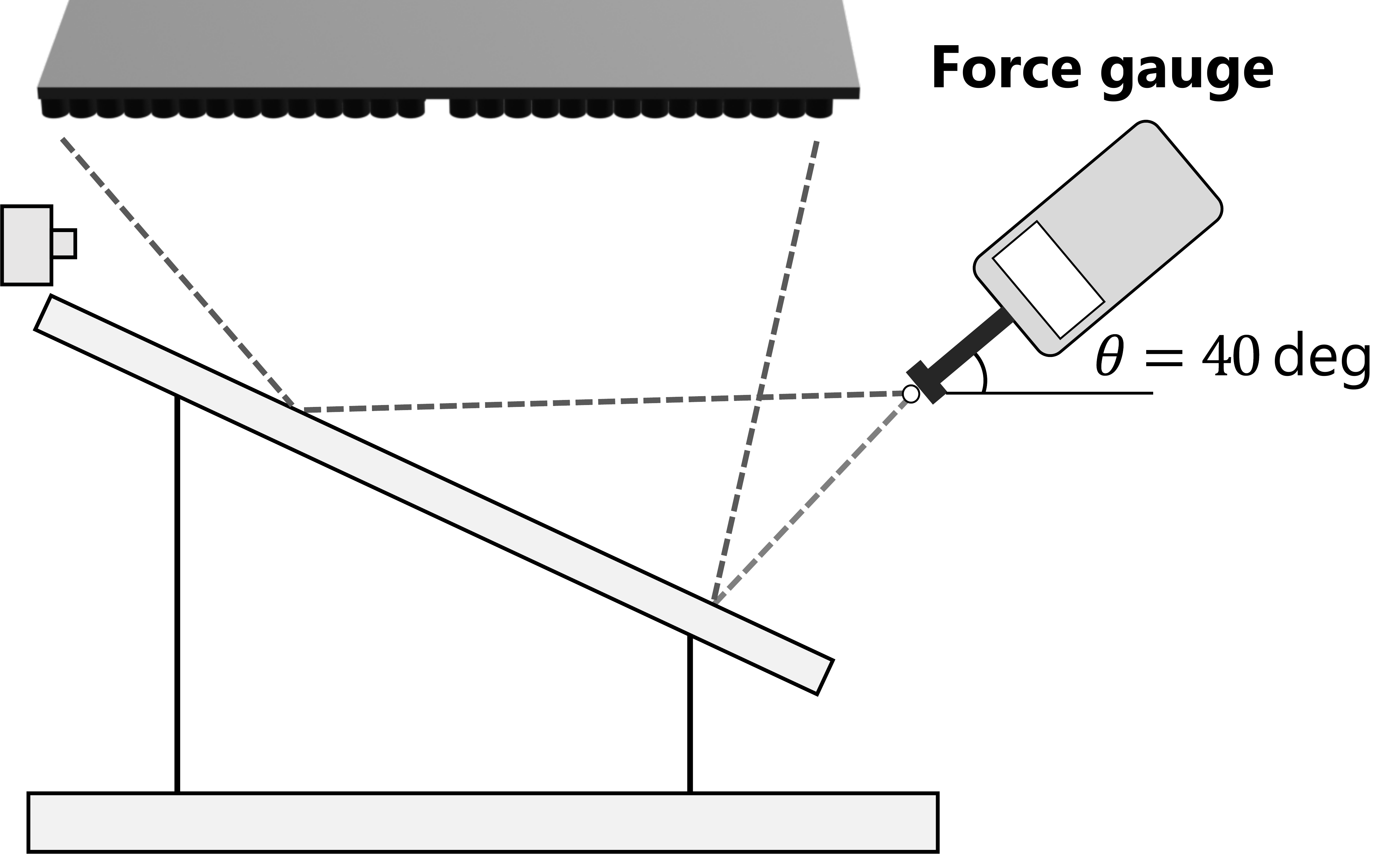}
    \caption{Setup for measuring radiation force. The tip of the force gauge to which a 1.5~cm diameter acrylic disk was attached was placed at the focal point. The force gauge was tilted 40 deg so that this disk opposed the propagation direction of the ultrasound wave. }
    \label{fig:Fig/Equipment/SetupMeasureForce.pdf}
\end{figure}

\begin{figure}[t]
    \centering
    \includegraphics[width=0.49\columnwidth]{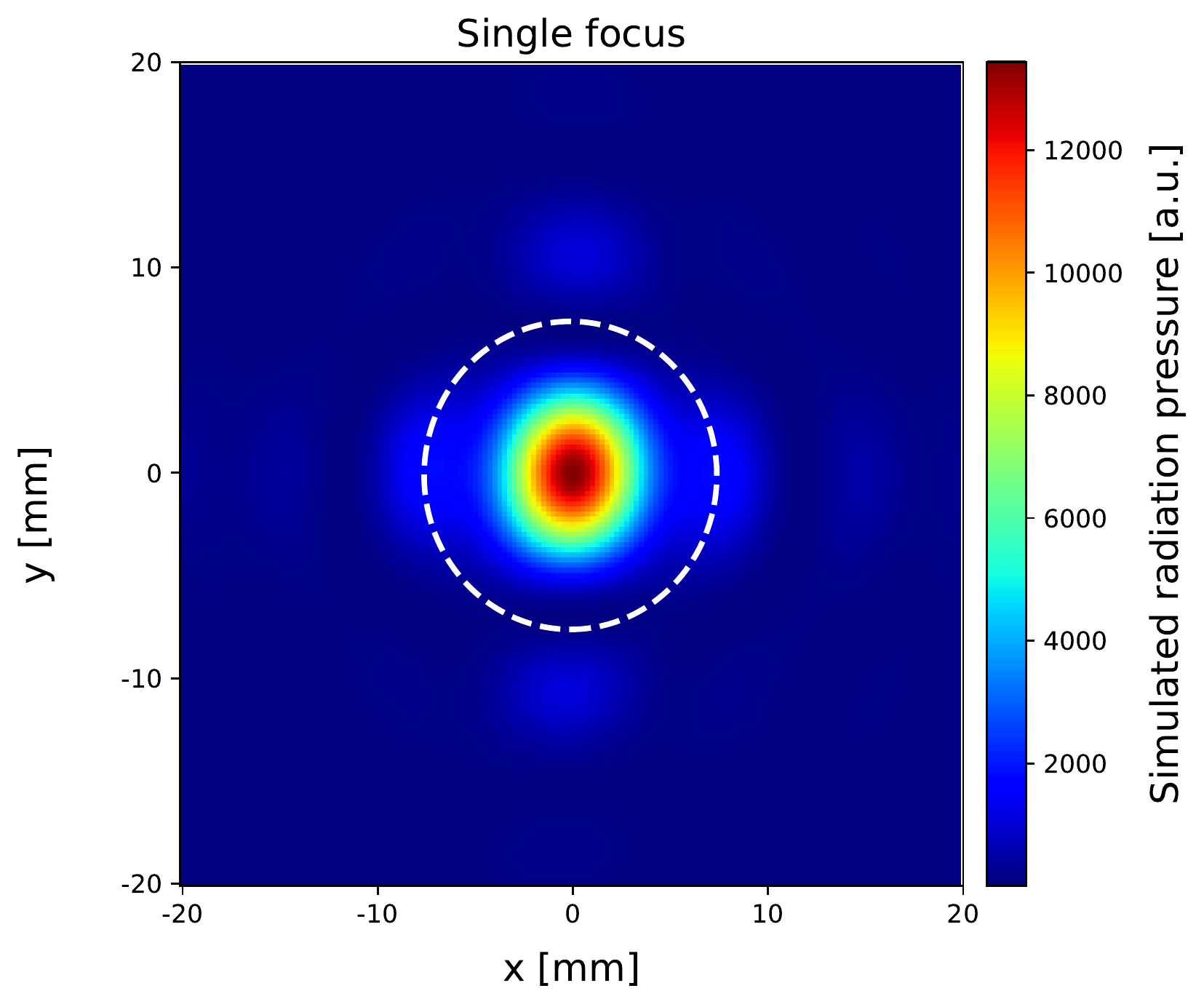}
    \includegraphics[width=0.49\columnwidth]{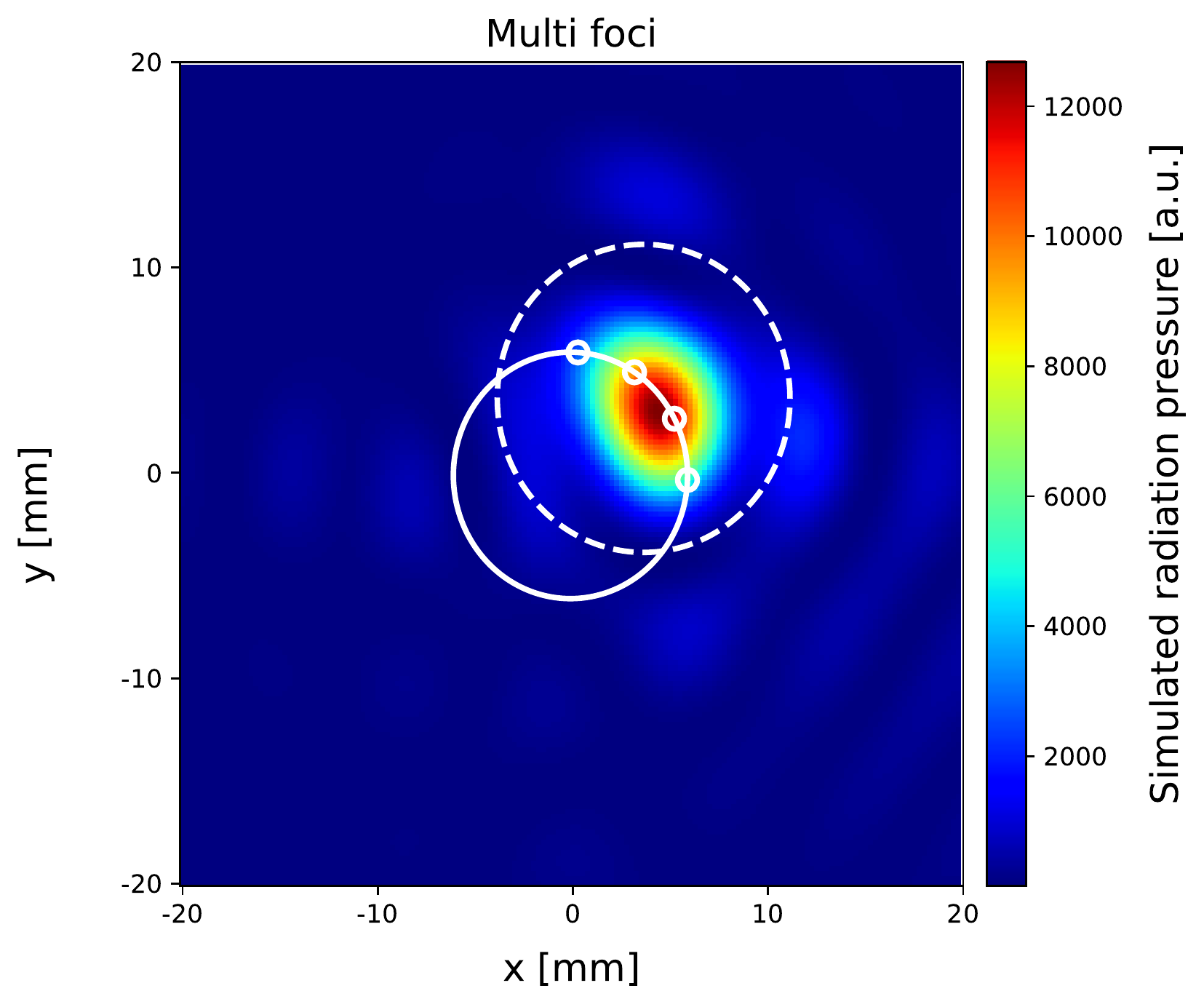}
    \caption{Simulated radiation pressure distribution of single focus and multi foci. The amplitude of each transducer was set to 1 a.u. The dotted white circle with a diameter of 1.5~cm indicates the area for measuring the radiation force. The continuous large and small white circle in LM-M simulation indicates the 6 mm radius LM trajectory and each focus position, respectively.}
\label{fig:Fig/Equipment/SimulatedRadiationPressure_simple.pdf}
\end{figure}

\begin{figure}[t]
    \centering
    \includegraphics[width=1\columnwidth]{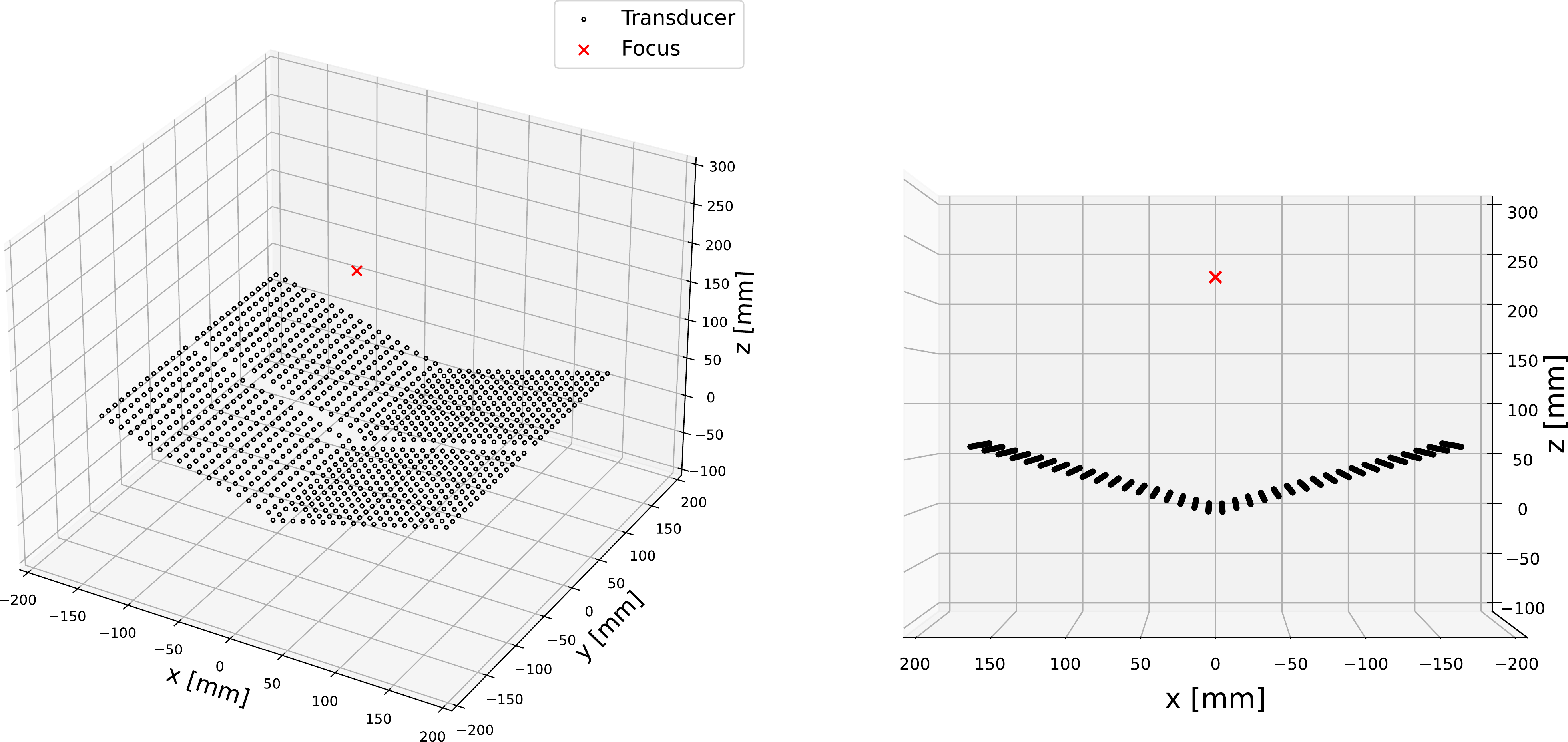}
    \caption{AUTD setup for the simulation.}
    \label{fig:Fig/Equipment/SetupAUTD.pdf}
\end{figure}

We measured the radiation force of the focus presented by the system at the center (0, 30, 30) mm, which was 0.022~N. The radiation force of LM-M with $\StimulusRadius = 6$ mm (four foci) was also the same. The LM-M was presented at the center, and the foci movement was stopped during the measurement ($\myTime = 0$). The radiation force was not varied when the focus and foci position shifted 6 mm around the center.

Fig.~\ref{fig:Fig/Equipment/SetupMeasureForce.pdf} shows the measurement setup. In this experiment, the tip of a force gauge, to which a 1.5~cm diameter acrylic disk was attached, was placed at the focal point. This force gauge (KYOWA LTS-50GA and WGA-900) can measure forces up to 0.5~N with a resolution of 0.001~N. The force gauge was tilted by 40 deg so that this disk opposes the propagation direction of the ultrasound wave. The size of the acrylic disk was larger than the created focus size to avoid underestimation of the radiation force. The measurement range of the acrylic disk is superimposed on the preliminary simulated focus (Fig.~\ref{fig:Fig/Equipment/SimulatedRadiationPressure_simple.pdf}) as a dotted white circle.

\begin{figure}[!t]
    \centering
    \includegraphics[width=0.8\columnwidth]{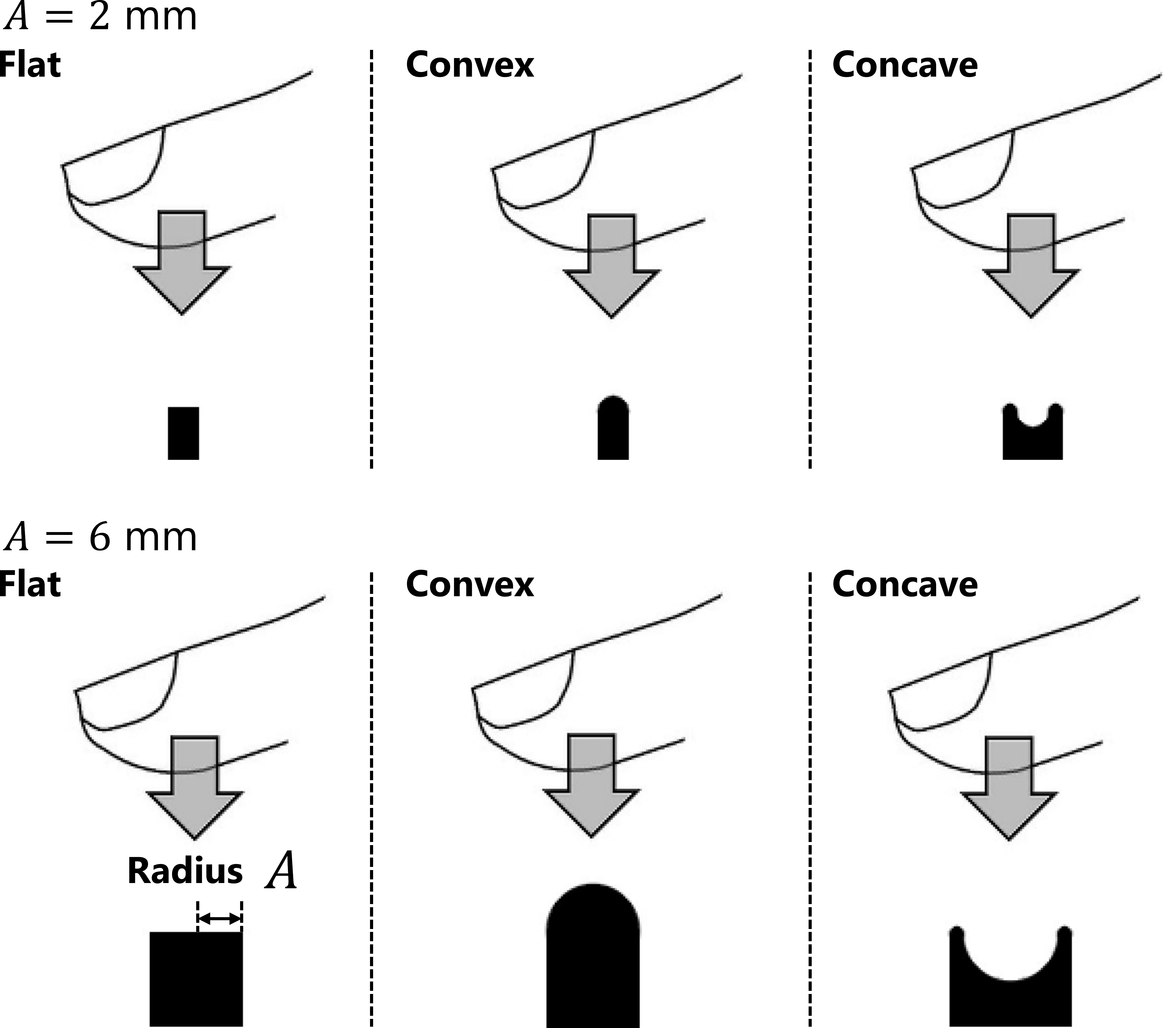}%
    \caption{Example image to evaluate the perceived curvature ($\StimulusRadius = 2,6$~mm). The object's radius was changed according to the radius of the presented LM stimulus $\StimulusRadius$. For one stimulus condition, the image of flat, convex, and concave was sequentially presented in random order. Participants reported the perceptual similarity between the perceived curvature and the image.}
    \label{fig:Fig/Ex1/ImageStimulusShape.pdf}
\end{figure}

\begin{figure*}[!t]
    \centering
    \subfloat[Evaluated vibration.]{\includegraphics[width=0.9\columnwidth]{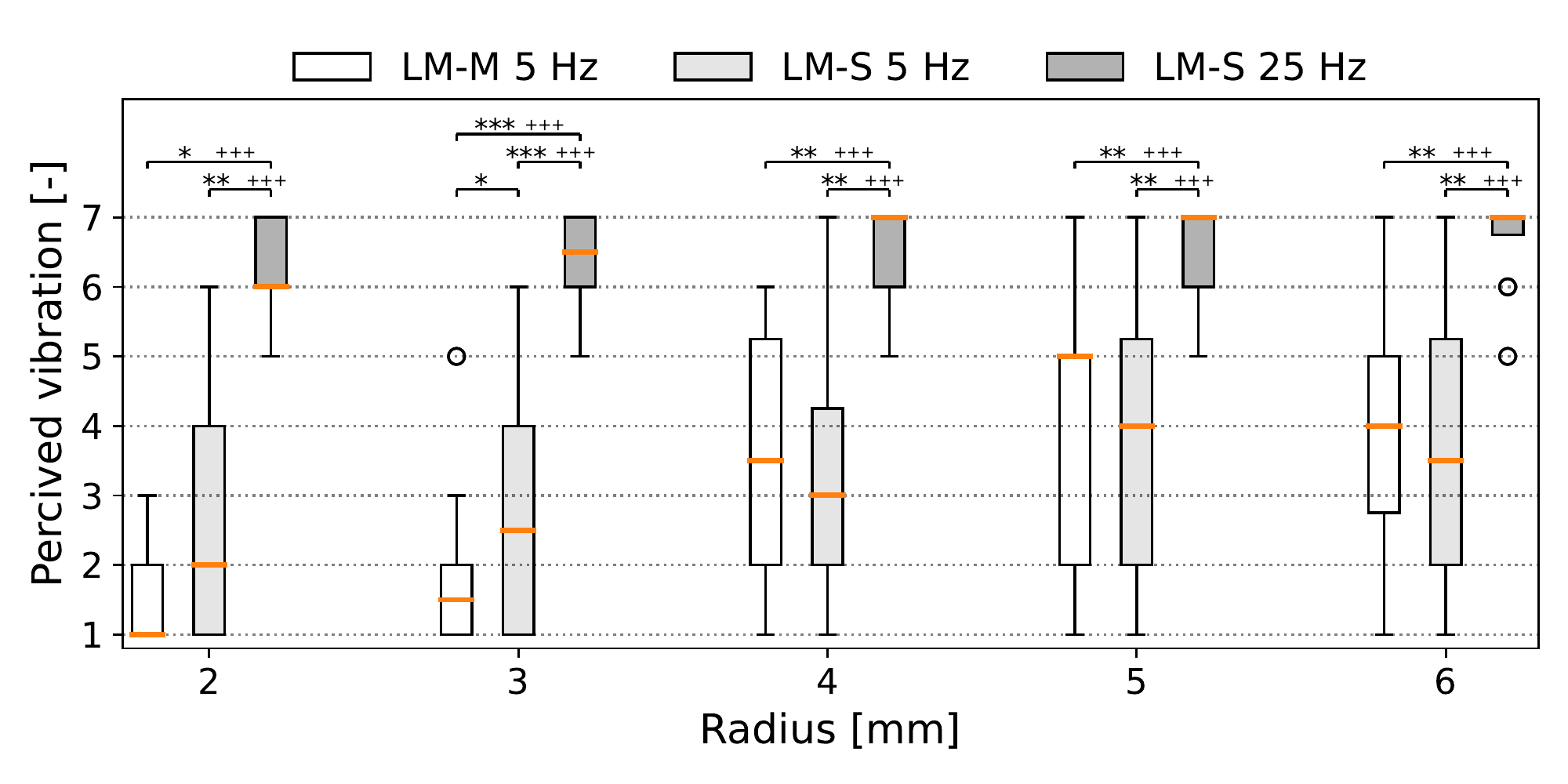}%
    \label{fig:Fig/Ex1/Vib_All.pdf}}
    \hfil
    \subfloat[Evaluated movement.]{\includegraphics[width=0.9\columnwidth]{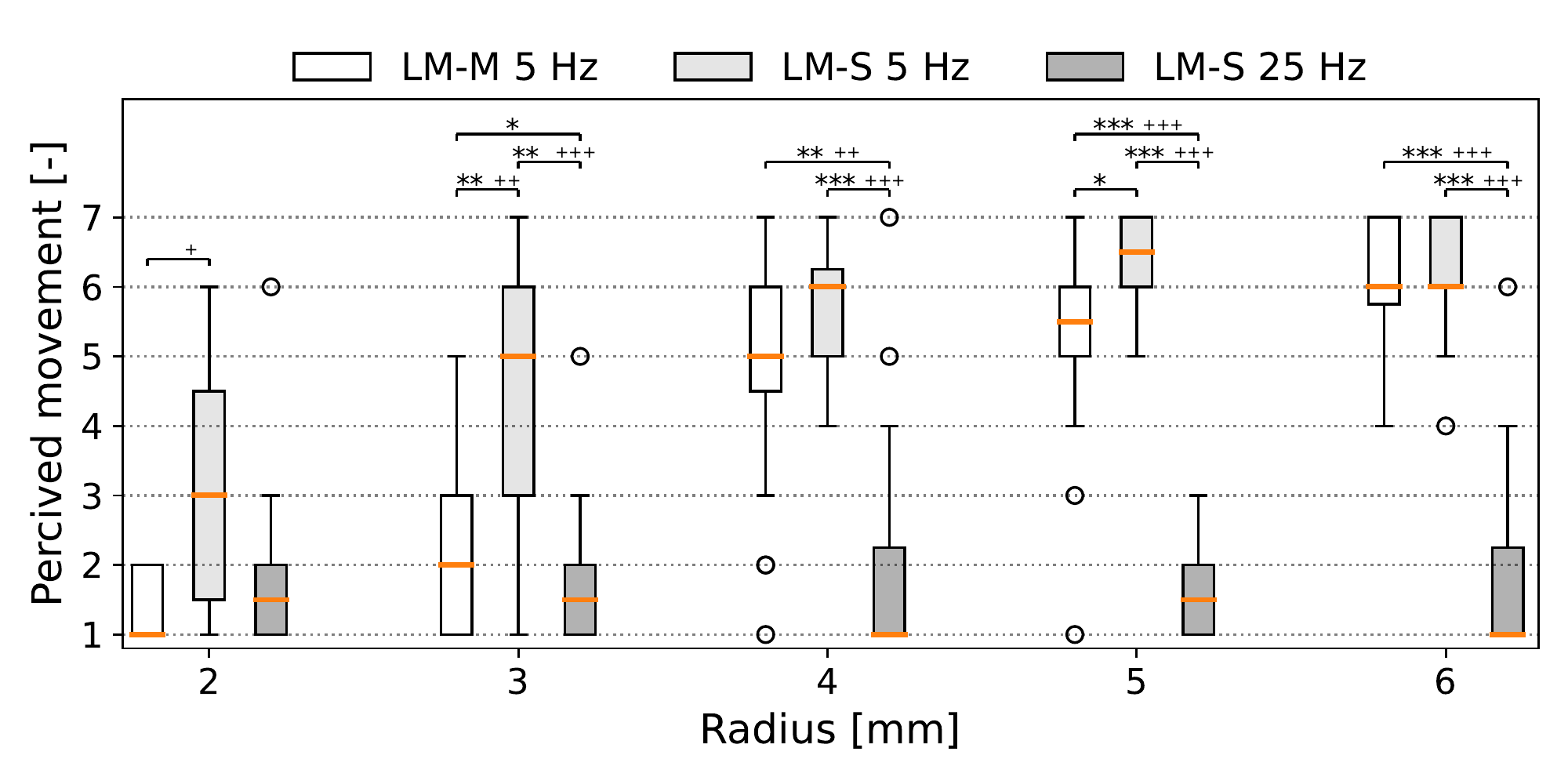}%
    \label{fig:Fig/Ex1/Movement_All.pdf}}
    \caption{Evaluated perceptual stationarity of LM stimulus on a finger pad in experiment 1. Participants evaluated the perceived intensity of the vibratory sensation and the focal movement sensation of the LM stimulus with a seven-point Likert scale.}
    \label{fig:Ex1 Evaluated tactile sensation}
\end{figure*}

\begin{figure}[!t]
    \centering
    \subfloat[LM-M at 5 Hz.]{\includegraphics[width=0.9\columnwidth]{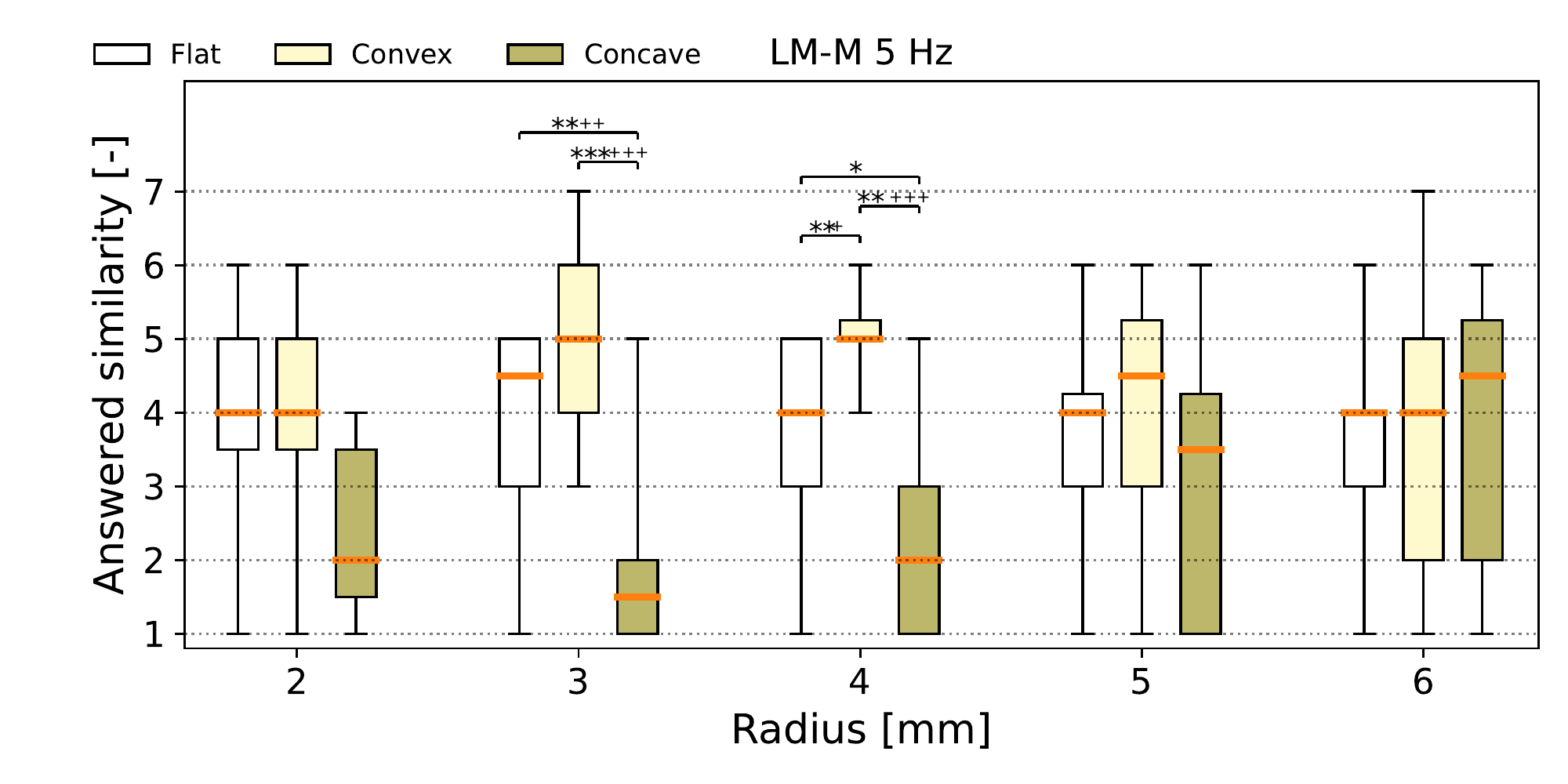}%
    \label{fig:Fig/Ex1/Shape_st_multi.pdf}}
    \hfil
    \subfloat[LM-S at 5 Hz.]{\includegraphics[width=0.9\columnwidth]{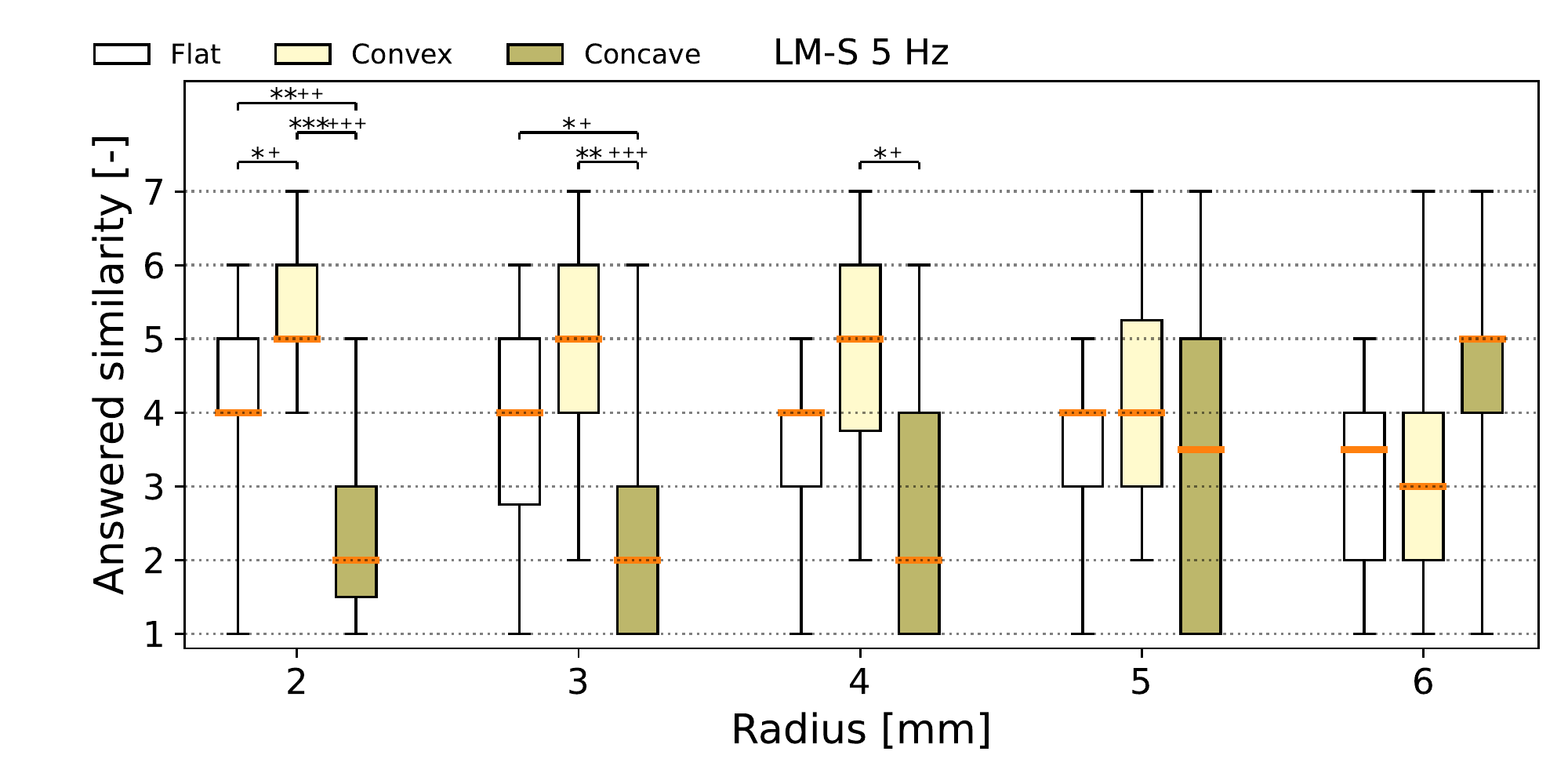}%
    \label{fig:Fig/Ex1/Shape_st_single.pdf}}
    \hfill
    \subfloat[LM-S at 25 Hz.]{\includegraphics[width=0.9\columnwidth]{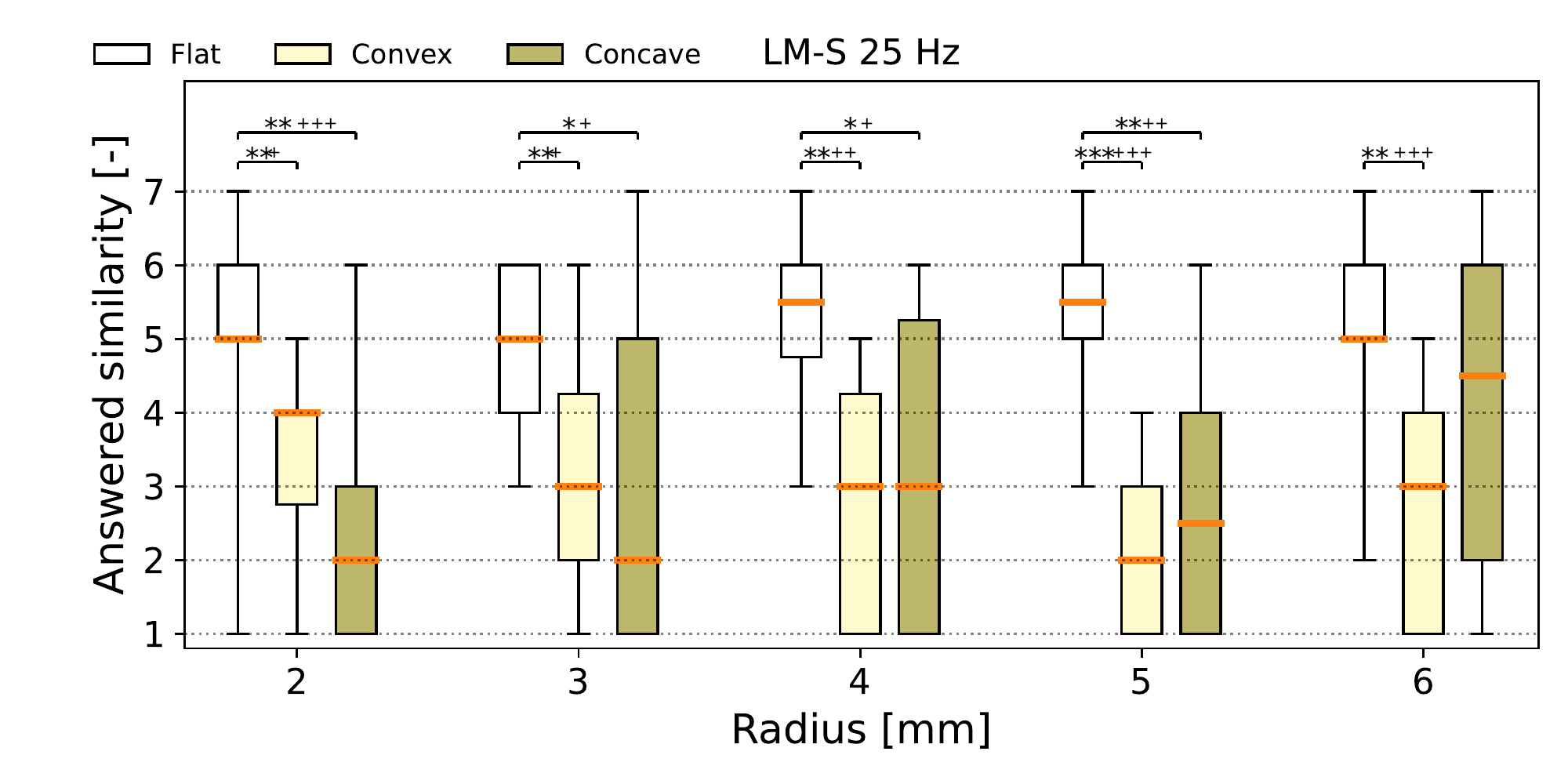}%
    \label{fig:Fig/Ex1/Shape_vib.pdf}}
    \caption{Evaluated perceived curvature in experiment 1. The reference images with flat, convex, and concave was presented, and the participants answered perceptual similarity between the perceived tactile shape (curvature) and the image with a seven-point Likert scale.}
    \label{fig:Ex1 Evaluated tactile shape}
\end{figure}

\section{Experiment1: Stationarity and \\Surface Curvature}\label{sec:Experiment1: Tactile Sensation and Shape}
This experiment evaluated the intensity of vibratory and movement sensations in the LM and the perceived surface curvature rendered by the LM (i.e., flat, convex, or concave). 

\subsection{Stimulus Condition\label{sec: Ex1 Stimulus condition}}
In this experiment, we presented the LM-M (LM-multi foci) and LM-S (LM-single focus) stimuli at 5 Hz (as described in Section~\ref{sec: Stimulus Design}) to present static pressure sensation. The previous study showed that LM at 5 Hz evokes pressure sensation~\cite{morisaki2021non}. For comparison, an LM-S stimulus at 25 Hz was also presented since 25 Hz vibration mainly stimulates RA-I tactile receptor~\cite{bolanowski1988four}. The radii of LM stimuli $\StimulusRadius$ were 2, 3, 4, 5, and 6~mm. The motion step width $\LmStep$ of the LM stimulus at 5 Hz was as fine as 0.23~mm to elicit static pressure sensation~\cite{morisaki2021non}. Moreover, the step $\LmStep$ at 25 Hz was 4~mm to avoid exceeding the AUTD update limits (1 kHz)~\cite{suzuki2021autd3}. For the 5 Hz LM-M stimuli, the number of simultaneously presented foci $\FocusNumTotal$ was four, and their placement interval $\MultiSpace$ was empirically set to 3~mm. All stimuli were presented in random order. Each participant underwent two sets of experiments. Therefore, 30 experimental trials were conducted (i.e., 3 different LM stimuli $\times$ 5 stimulus radii $\times$ 2 sets = 30 experimental trials).

\subsection{Procedure\label{sec: Ex1 Procedure}}
Eight males (24--31 age) and two females (24 and 28 age) participated in this experiment. All participants participated in Experiment 2 (described in Section~\ref{sec: Perceived Size}). The three males and the three females also participated in Experiment 3 (described in Section~\ref{sec:  Equivalent Physical Stimulus}). All participants have experienced ultrasound midair haptic systems.

The experimental equipment was a visuo-tactile display (Fig~\ref{fig:Fig/Equipment/ExperimentalSetupActual.pdf} and Section~\ref{sec: Experimental Equipment}). Participants were instructed to place their index fingertips of the right hand on the presented midair image marker. The tactile stimulus was always presented while the fingertip was touching the marker.

First, to evaluate the tactile sensation of the presented stimulus, the participants answered the following two questions with a seven-point Likert scale:
\begin{description}
\item[Q1.] How intensely did you perceive a vibratory sensation in the presented stimulus?
\item[Q2.] How intensely did you perceive the movement of the stimulus position?
\end{description}
Participants were instructed to answer 1 if they perceived no vibration or movement. In Q2, we evaluated whether the participants noticed the circular focus movement of the LM. 

Second, the participants evaluated the curvature rendered by the LM stimulus on their finger pads. In this experiment, we provided three typical shapes as references (i.e., flat, convex, and concave). Three images corresponding to the three shapes (Fig.~\ref{fig:Fig/Ex1/ImageStimulusShape.pdf}) were presented to the participants as reference images. 

To evaluate the perceived curvature, the participants responded to Q3 with a seven-point Likert scale. 
\begin{description}
\item[Q3.] Does the stimulus shape perceived at your finger pad match the situation illustrated in the reference images?
\end{description}
For one stimulus condition, flat, convex, and concave reference images (Fig.~\ref{fig:Fig/Ex1/ImageStimulusShape.pdf}) were presented successively in random order. Participants independently reported perceptual similarity to each reference image (i.e., flat, convex, and concave). We varied the radius of the illustrated object in the reference images to match that of LM stimulus $\StimulusRadius$. The convex and concave part was a circle with a radius $\StimulusRadius$.

Participants were instructed to ignore differences in the perceived size between the image and tactile stimulus to evaluate only the similarity of the perceived curvature (i.e., flat, convex, and concave). The overall size of the finger sketch, which was drawn in the reference image, was adjusted so that its nail size matches the average Japanese adult nail length (13.6~mm)~\cite{AIRC2011HandSize}. 

%\Add{They were also instructed to respond with a low score for all shapes if the stimulus was not perceived as flat, convex, or concave.} 

\subsection{Results and Analysis}
\subsubsection{Stationarity}
\newcommand{\data}{v}
\newcommand{\DataSevenFive}{\data^\mathrm{75}}
\newcommand{\DataTwoFive}{\data^\mathrm{25}}
\newcommand{\IQR}{IQR}
\newcommand{\pValue}{p}
Box-and-whisker plots of the evaluated vibratory sensations (answers to Q1) are shown in Fig.~\ref{fig:Fig/Ex1/Vib_All.pdf}. The evaluated movement sensation (answers to Q2) is also shown in Fig.~\ref{fig:Fig/Ex1/Movement_All.pdf}. If the data value $\data$ satisfies the following conditions, the data are treated as an outlier:
\begin{IEEEeqnarray}{rCl}
\left\{
\begin{array}{l}
\data \leq \DataTwoFive - 1.5 \times \IQR,\label{eq:outlier}\\
\data \geq \DataSevenFive + 1.5 \times \IQR,
\end{array}
\right.
\end{IEEEeqnarray}
where $\DataTwoFive$ and $\DataSevenFive$ are the 25-percentile value and 75-percentile value, respectively, and $IQR$ is the interquartile range. Outliers were plotted as white dots in the graphs. As seven participants could not perceive the LM-M stimulus with $\StimulusRadius = 2$~mm, their answers were excluded. In total, 13 data of the LM-M with $\StimulusRadius = 2$~mm were excluded from each graph. The results showed that the lowest median value of the vibratory sensation score was 1, and the condition was LM-M at 5 Hz with $\StimulusRadius = 2$~mm. At the condition, the movement sensation was 1 and the lowest.

Shapiro-Wilk test indicated that the perceived vibration and movement data were not normally distributed ($\pValue < 0.005$). We conducted the Wilcoxon signed-rank test with Bonferroni correction between the LM type for each $\StimulusRadius$. The results of the LM-M with $\StimulusRadius = 2$~mm were excluded from the analysis. Fig.~\ref{fig:Ex1 Evaluated tactile sensation} shows the pairs with significant differences as "*", "**", and "***" for $\pValue < 0.05$, $< 0.005$, and $< 0.0005$, respectively. The test results showed that with all $\StimulusRadius$, the vibratory sensation of the LM-S at 25 Hz was significantly higher than that of the other LM stimuli ($\pValue < 0.0005$). With $\StimulusRadius = 3, 5$~mm, the movement sensation of the LM-M was significantly lower than that of the LM-S at 5 Hz ($\pValue < 0.05$). We conducted the Statistical power analysis of t-test which is a parametric test corresponding to the Wilcoxon signed-rank test. The statistical power is shown in Fig.~\ref{fig:Ex1 Evaluated tactile sensation}. “+”, “++”, and “+++” indicate the statistical power greater than 0.8, 0.98, and 0.998, respectively. We also conducted the Friedman test with Bonferroni correction with the LM type and $\StimulusRadius$. The LM type had a significant effect on both the vibration and the movement sensation and the $\StimulusRadius$ had it only on the movement sensation ($\pValue < 0.0005$).

\subsubsection{Surface Curvature}
Fig~\ref{fig:Ex1 Evaluated tactile shape} show the evaluated tactile shape (answers to Q3). 13 data of LM-M with $\StimulusRadius = 2$~mm were excluded (Section~\ref{sec: Ex1 Procedure}). The highest median for the convex score was 5, and the conditions were LM-M with $\StimulusRadius = 3,4$~mm and LM-S at 5 Hz with $\StimulusRadius = 2, 3, 4$~mm. At these conditions, the flat and convex score was lower than 4.5.

Shapiro-Wilk test indicated that the perceived curvature data was not normally distributed ($\pValue < 0.005$). We conducted the Wilcoxon signed-rank test with Bonferroni correction to compare the score between the shapes (i.e., flat, convex, concave) at each LM type. The test result showed that the convex scores were significantly higher than the concave and float scores ($\pValue < 0.05$) in the LM-M with $\StimulusRadius = 4$~mm and the LM-S at 5 Hz with $\StimulusRadius = 2$~mm. We conducted the Statistical power analysis of t-test. The statistical power is shown in Fig.~\ref{fig:Ex1 Evaluated tactile shape}. We also conducted the Friedman test with Bonferroni correction with the LM type and $\StimulusRadius$. The LM type had a significant effect on the flat and convex score and the $\StimulusRadius$ had it on the convex and concave score ($\pValue < 0.0005$).

\section{Experiment2: Perceived Size}\label{sec: Perceived Size}

\begin{figure}[!t]
    \centering
    \includegraphics[width=0.9\columnwidth]{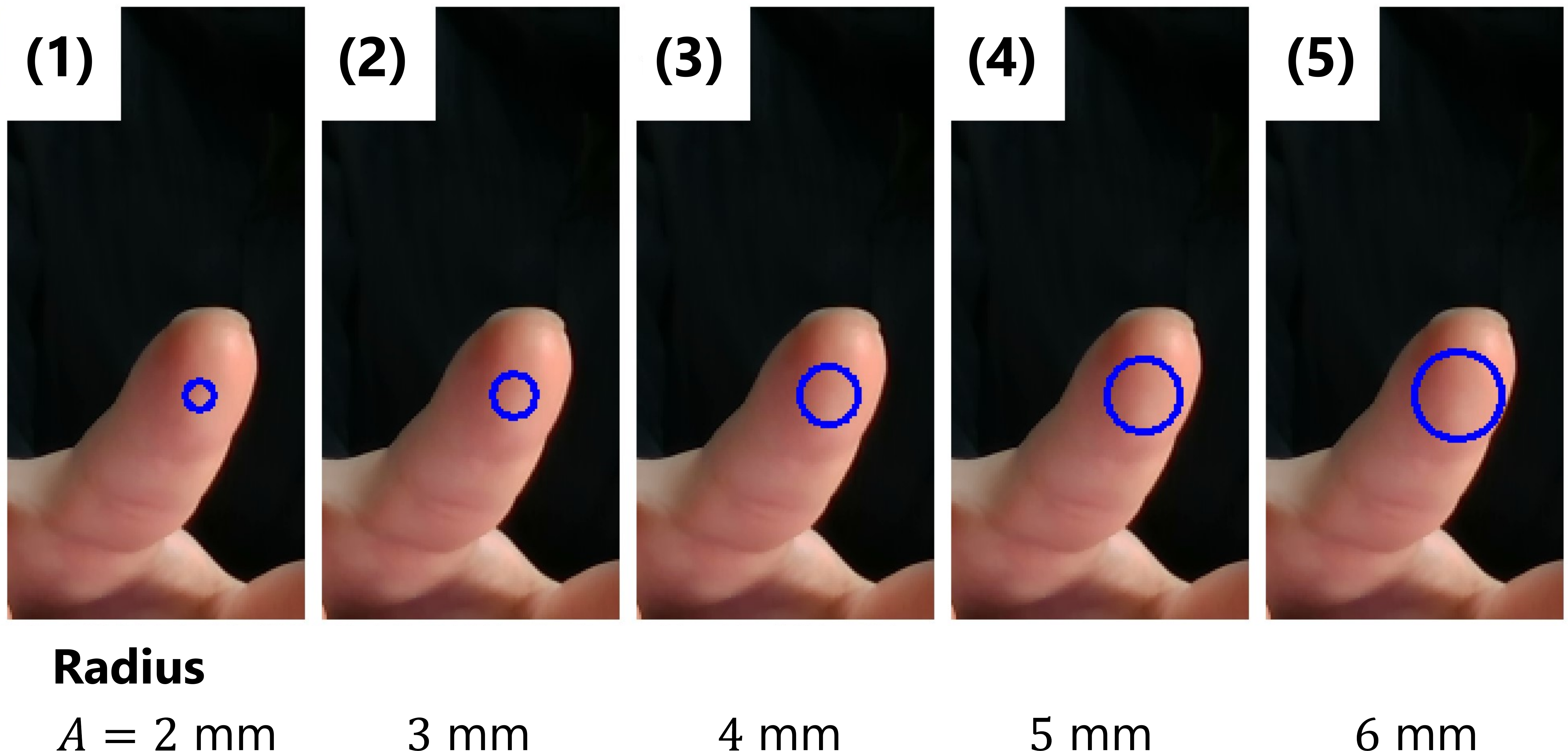}%
    \caption{Presented picture to evaluate the perceived size of the presented LM stimulus. The five pictures with different radii (2, 3, 4, 5, and 6~mm) were presented simultaneously. Participants selected one of these images showing the circle whose size matches the perceived haptic size.}
    \label{fig:Fig/Ex2/ImageStimulusSize.pdf}
\end{figure}

\begin{comment}
 \begin{figure*}
    \centering \includegraphics[width=1.9\columnwidth]{Fig/Ex2/Mat_ALL.pdf}%
    \caption{Confusion matrix of the stimulus size identification. The chance rate in this experiment was 0.2.}
    \label{fig:Fig/Ex2/Mat_ALL.pdf}
\end{figure*}   
\end{comment}

\begin{figure}
    \centering \includegraphics[width=0.9\columnwidth]{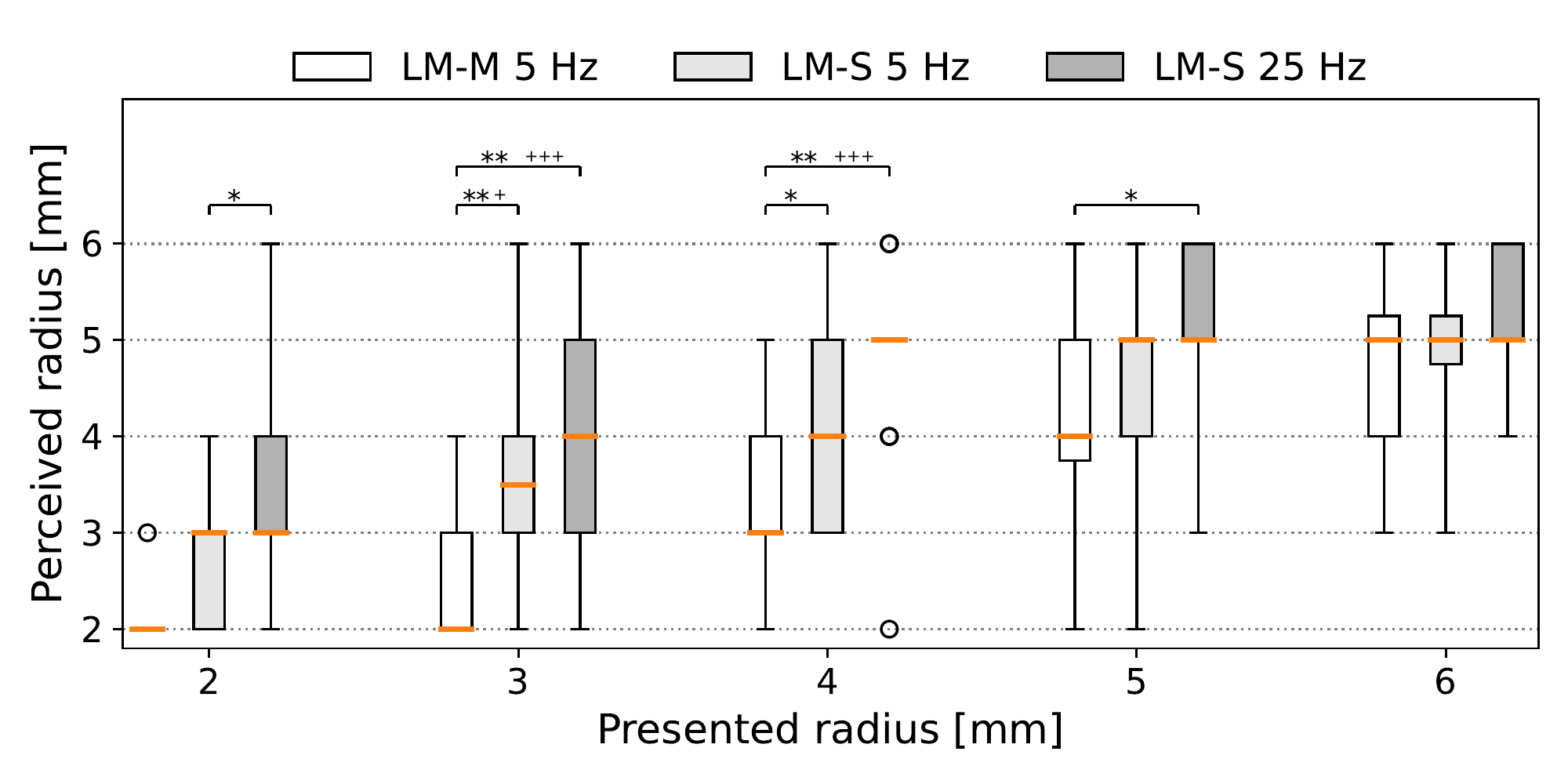}%
    \caption{Evaluation result of the perceived size of the circular LM stimulus.}
    \label{fig:Fig/Ex2/Analog_Size_Comp.pdf}
\end{figure}
In this experiment, we changed the radius of LM stimulus $\StimulusRadius$ and evaluated the perceived stimulus size.

\subsection{Procedure}
Eight males (24--31 age) and (24 and 28 age) two females participated in this experiment. 

The experimental setup was the same as that used in Experiment 1 (Fig.~\ref{fig:Fig/Equipment/ExperimentalSetupActual.pdf}). The tactile stimulus was always presented while the index fingertip of the right hand was touching the marker. The stimulus conditions were identical to those used in Experiment 1, which is explained in Section~\ref{sec: Ex1 Stimulus condition}. 30 experimental trials were conducted (i.e., 3 different LM stimuli $\times$ 5 stimulus radii $\times$ 2 sets = 30 experimental trials).

A real-time video of the participants' fingers was presented to them during the experiment. The screenshot of the video is shown in Fig.~\ref{fig:Fig/Ex2/ImageStimulusSize.pdf}. In this video, a blue circular image corresponding to the trajectory of the LM is superimposed on the finger pad. Participants selected one of the videos showing a circle whose size matched the perceived haptic size to evaluate the perceived size of the stimulus. 

The center of the circular image was changed in real-time to match the center of the presented LM stimulus $\CenterPosition$. The radii of the circular images were 2, 3, 4, 5, and 6~mm, which were the same as the radii of LM stimuli $\StimulusRadius$ used in this experiment. Five videos with different radii were simultaneously presented to the participant. This video was captured using an RGB camera built into the depth camera. 

\subsection{Results and Analysis}
Fig.~\ref{fig:Fig/Ex2/Analog_Size_Comp.pdf} shows the perceived stimulus sizes. The highest perceived stimulus radius was 5~mm, and the condition was LM-S at 5 and 25 Hz with $\StimulusRadius = 5$~mm and all LM stimuli with $\StimulusRadius = 6$~mm. The lowest radius was 2~mm, and the conditions were LM-M with $\StimulusRadius = 2, 3$~mm.

Shapiro-Wilk test indicated that the perceived size data were not normally distributed ($\pValue < 0.0005$). The Wilcoxon signed-rank test with Bonferroni correction showed that with $\StimulusRadius = 3, 4$~mm, the perceived radius of the LM-M was significantly lower than that of the LM-S at 5 and 25 Hz ($\pValue < 0.05$). We conducted the Statistical power analysis of t-test. The statistical power is shown in Fig.~\ref{fig:Fig/Ex2/Analog_Size_Comp.pdf}. We also conducted the Friedman test with Bonferroni correction with the $\StimulusRadius$ and the LM type. Both the $\StimulusRadius$ and the LM type had a significant effect on the perceived size ($\pValue < 0.0005$).

\section{Experiment3: Equivalent Physical Stimulus}\label{sec:  Equivalent Physical Stimulus}
\begin{figure}
    \centering \includegraphics[width=0.85\columnwidth]{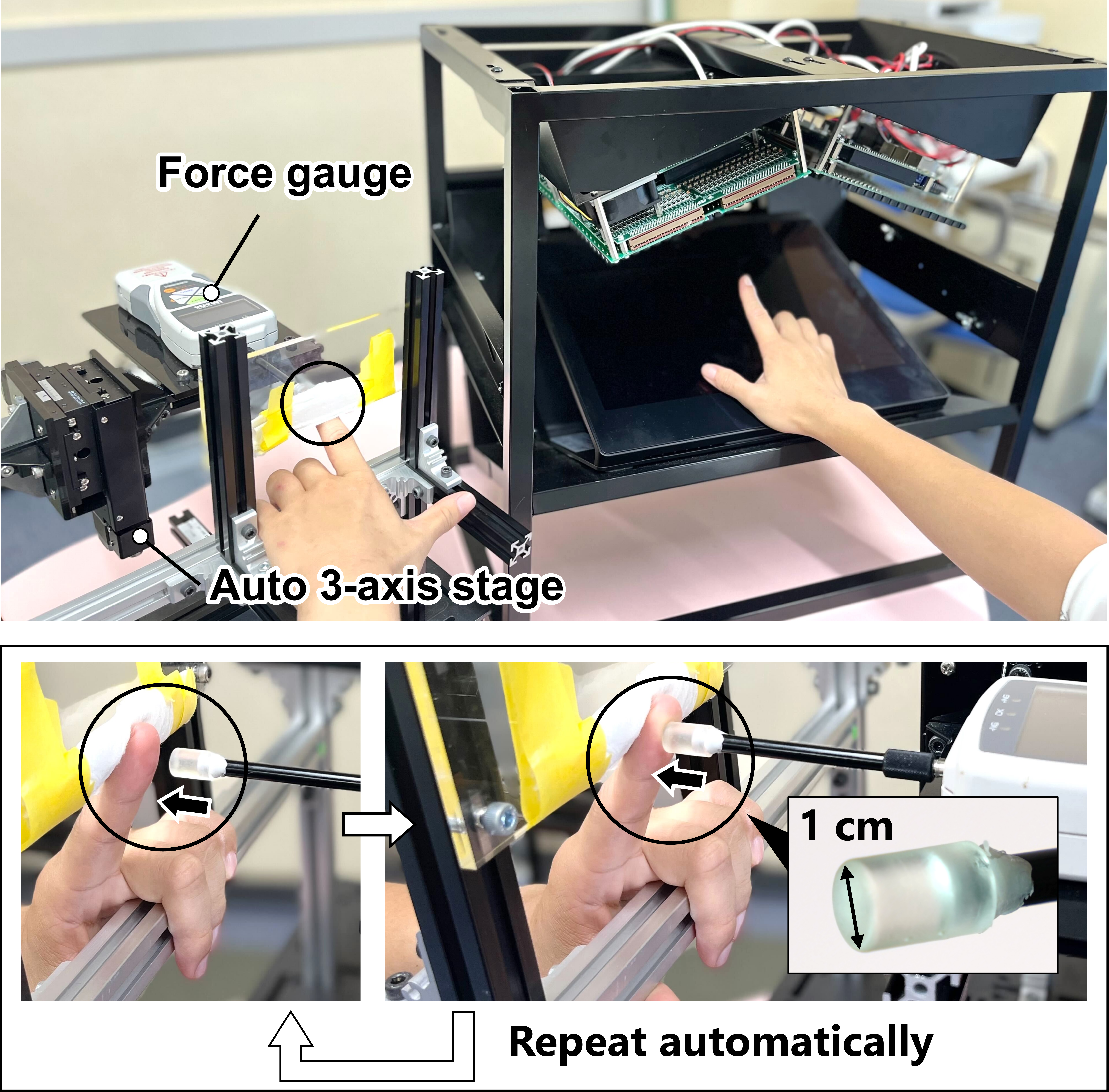}%
    \caption{Setup to evaluate the perceived force. A force gauge was pressed against the finger pad of the left hand, and the LM stimulus was presented to the right finger. The pushing depth was automatically controlled by the 3-axis stage. These stimuli were terminated after 2 s and automatically repeated. Participants compared the pushing force with the LM stimulus and orally reported the comparison results.}
    \label{fig:Fig/Ex3/ForceGaugeSetup.pdf}
\end{figure}

This experiment investigated physical force which is equivalent to the pressure sensation evoked by LM. Physical force was presented by pushing a force gauge against the finger pad.

\subsection{Setup and Stimulus}
Fig.~\ref{fig:Fig/Ex3/ForceGaugeSetup.pdf} illustrated the experimental setup. In this experiment, we used a force gauge whose z-position was automatically controlled by a 3-axis stage (QT-AMM3 and ALS-7013-G1MR, CHUO PRECISION INDUSTRIAL Co., Ltd.) and the visual-haptic system (Fig.~\ref{fig:Fig/Equipment/ExperimentalSetupActual.pdf}) used in the other experiments. This force gauge (IMADA ZTS-50N) can measure forces up to 50~N with a resolution of 0.01~N. The stimulus condition was the same as that used in Experiment 1, which is explained in Section~\ref{sec: Ex1 Stimulus condition}. 30 experimental trials were conducted (3 LM types $\times$ 5 stimulus radii $\StimulusRadius$ $\times$ 2 sets = 30 trials).

\subsection{Procedure}
Eight males (23--28 age) and two females (24 and 28 age) participated in this experiment. 

Participants were instructed to place their index fingers of their right hands on the marker presented by the midair image display. Participants were also instructed to place their index fingers of their left hands such that the finger pad faced the tip of the force gauge. At this point, the force gauge did not touch the finger pad. The force gauge was fixed in midair in a horizontal orientation (Fig.~\ref{fig:Fig/Ex3/ForceGaugeSetup.pdf}). Participants grasped the aluminum handle and fixed their finger position by placing it in front of an acrylic auxiliary plate. A plastic cylinder with a radius of 1~cm was attached to the tip of the force gauge. The basal plane of the cylinder was beveled to 1~mm so that the participants did not perceive its edges. Participants wore headphones and listened to white noise during the experiment to avoid hearing the driving noise of the AUTD.

A force gauge was pressed against the finger pad of the participant by moving along the z-axis. After the force gauge reached the specified position (the initial pushing depth was 4~mm), an LM stimulus was presented to the finger pad of the right hand. After 2 s, the LM stimulus was stopped, and the force gauge returned to its initial position. The force gauge immediately started pushing again, and the LM stimulus was presented again. This 2 s tactile stimulation was repeated automatically. In this experimental loop, participants compared the physical pushing force with the LM stimulus and orally reported the results. Based on the participants' answers, we changed the pushing depth of the force gauge such that the perceived intensity of the two stimuli is the same. For example, pushing depth in the 2nd stimulus was shortened to weaken the pushing force if the participant answered that the pushing force was stronger than the LM stimulus in the 1st stimulus. The force gauge kept pushing the finger pad and recorded the pushing force for 2 s when the participants reported that the intensities of the two stimuli were the same. The median value of the pushing force time series data was finally adopted as the measured force. After the measurement, the stimulus conditions were changed, and the same procedure was repeated. The adjustment resolution of the pushing depth is 0.25~mm and the speed of the force gauge was 5~mm/s. The maximum number of pushing depth adjustments was 20, and all participants completed the experiment within 30 min.

In the stimulus comparison, we instructed the participants to ignore the perception at the moment when the LM stimulus and the pushing force were presented to assess the steady-state perceived intensity of the LM stimulus. 

\subsection{Results and Analysis}
\begin{figure}[!t]
    \centering
    \includegraphics[width=0.9\columnwidth]{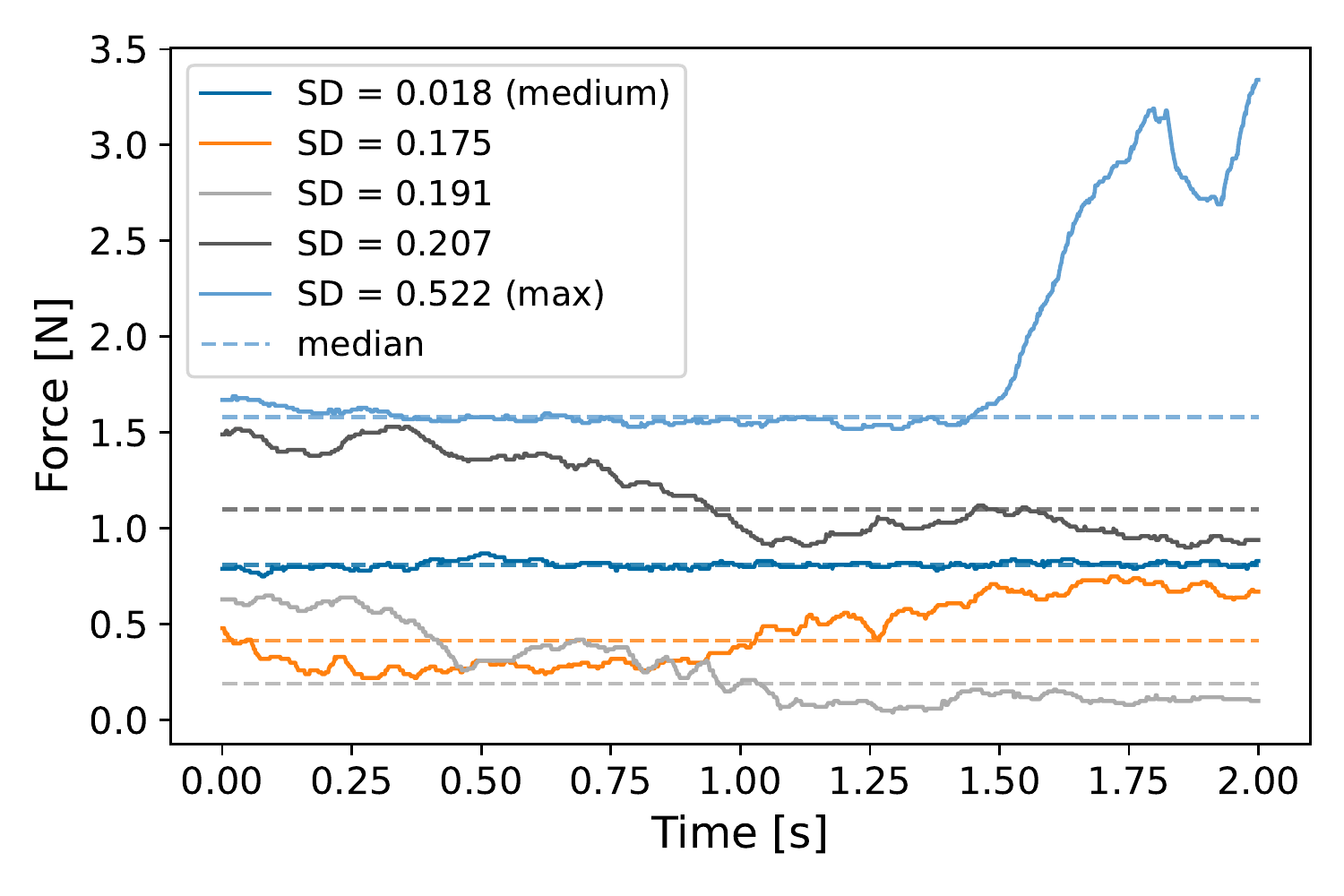}%
    \caption{Time series data of measured force. We calculated the standard deviation (SD) of each recorded time series data. The data with SD greater than 0.1 and with a medium SD (0.018) was plotted. The median value of these plotted time series force was also plotted as a dotted line.}
    \label{fig:Fig/Ex3/TimeSeriseForce.pdf}
\end{figure}

\begin{figure}[!t]
    \centering
    \includegraphics[width=0.9\columnwidth]{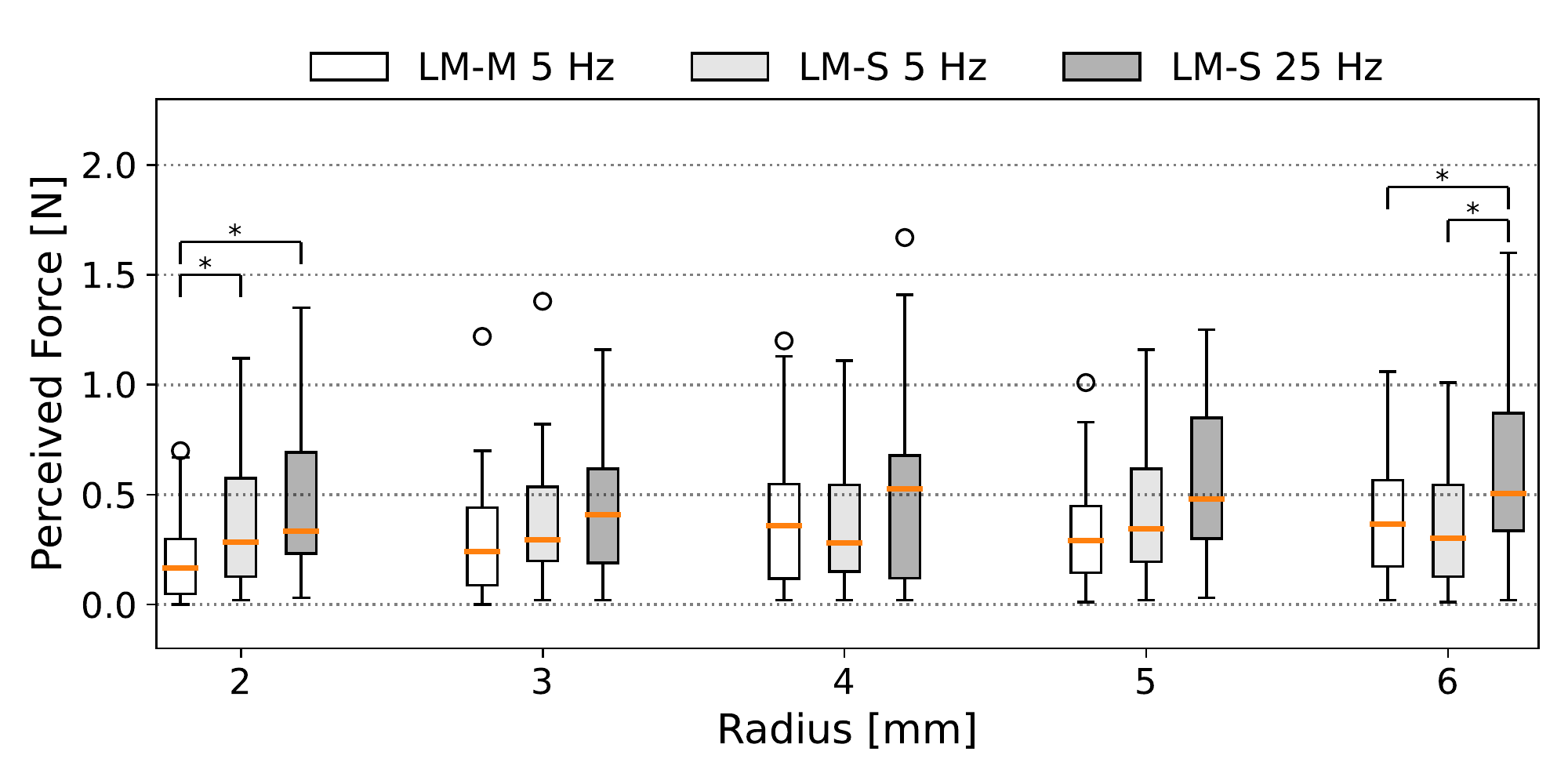}%
    \caption{Physically static pushing force perceptually equal to the intensity of the LM stimulus.}
    \label{fig:Fig/Ex3/PerceivedForce.pdf}
\end{figure}

The median value of the measured force time series data was adopted as the participant's answer. The maximum standard deviation (SD) of the time series data, median value, and minimum values were 0.522, 0.186, and 0, respectively. The maximum, median, and minimum values were answered by different participants. Fig.~\ref{fig:Fig/Ex3/TimeSeriseForce.pdf} shows the times series data with SD greater than 0.1 N and with the medium SD (0.018) was plotted. The median value of these plotted time series force was also plotted as a dotted line.

Fig.~\ref{fig:Fig/Ex3/PerceivedForce.pdf} shows the box-and-whisker plots of the pushing forces. Outliers were calculated using eq.~\ref{eq:outlier}, and are plotted as white dots. One participant was unable to perceive the LM at 5 Hz with $\StimulusRadius = 2$~mm; thus, this value was plotted as 0~N. The forces lower than 0.01~N, which is the lowest measurable force of the force gauge, were also plotted as 0~N. The results showed that the highest median value of the perceived force was 0.53~N, and the stimulus condition was LM-S at 25 Hz with $\StimulusRadius = 4$~mm. The lowest median value was 0.16~N, and the condition was LM-M at 5 Hz with $\StimulusRadius=2$~mm.

Shapiro-Wilk test indicated that the perceived force data were not normally distributed ($\pValue < 0.0005$). The Wilcoxon signed-rank test with Bonferroni correction showed that there are no significant difference in the perceived force expert four pairs ($\pValue > 0.05$, shown in Fig.~\ref{fig:Fig/Ex3/PerceivedForce.pdf}). The calculated statistical power is also shown in Fig.~\ref{fig:Fig/Ex3/PerceivedForce.pdf}. We conducted the Friedman test with Bonferroni correction with the $\StimulusRadius$ and the LM type. The $\StimulusRadius$ ($\pValue < 0.05$) and the LM type ($\pValue < 0.0005$) had a significant effect on the perceived force.

\section{Discussion}
\begin{figure*}[!t]
    \centering
    \includegraphics[width=1.9\columnwidth]{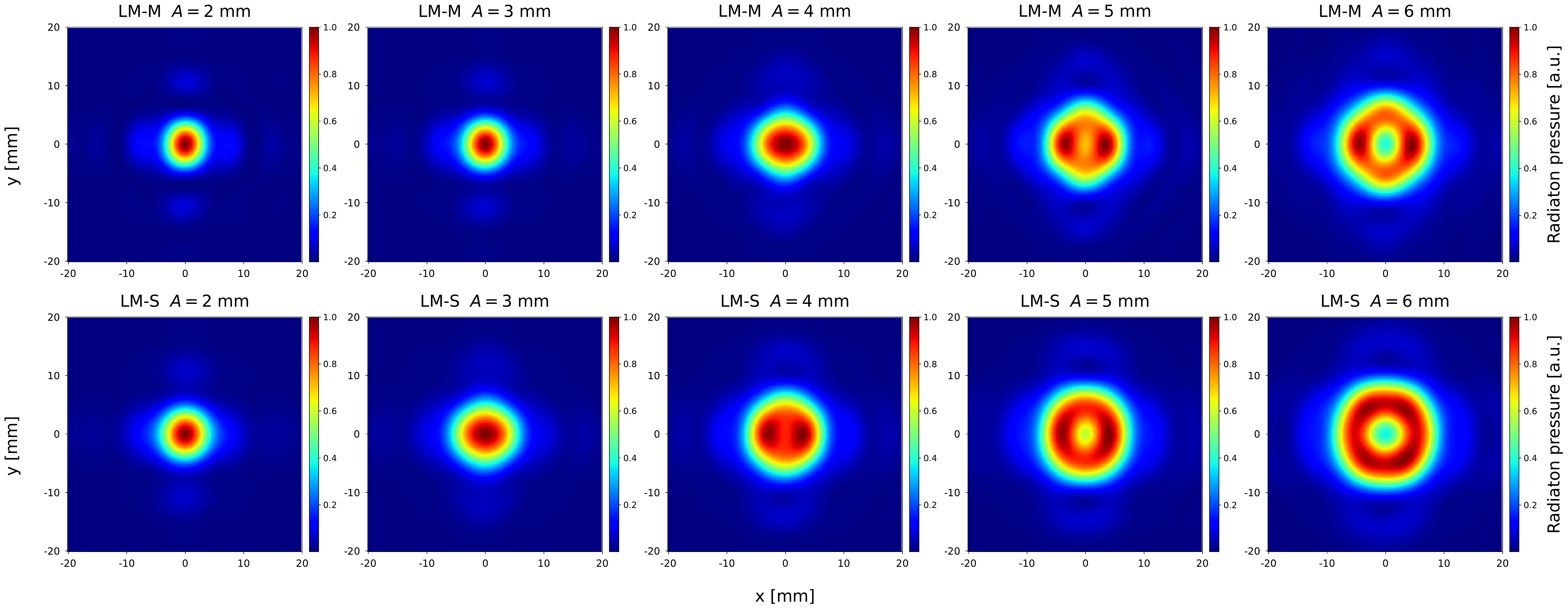}%
    \caption{Simulated time-averaged radiation pressure distribution. These values were normalized.}
    \label{fig:Fig/Discussion/RadiationPressureDistribu.pdf}
\end{figure*}
\begin{figure}[!t]
    \centering
    \includegraphics[width=0.9\columnwidth]{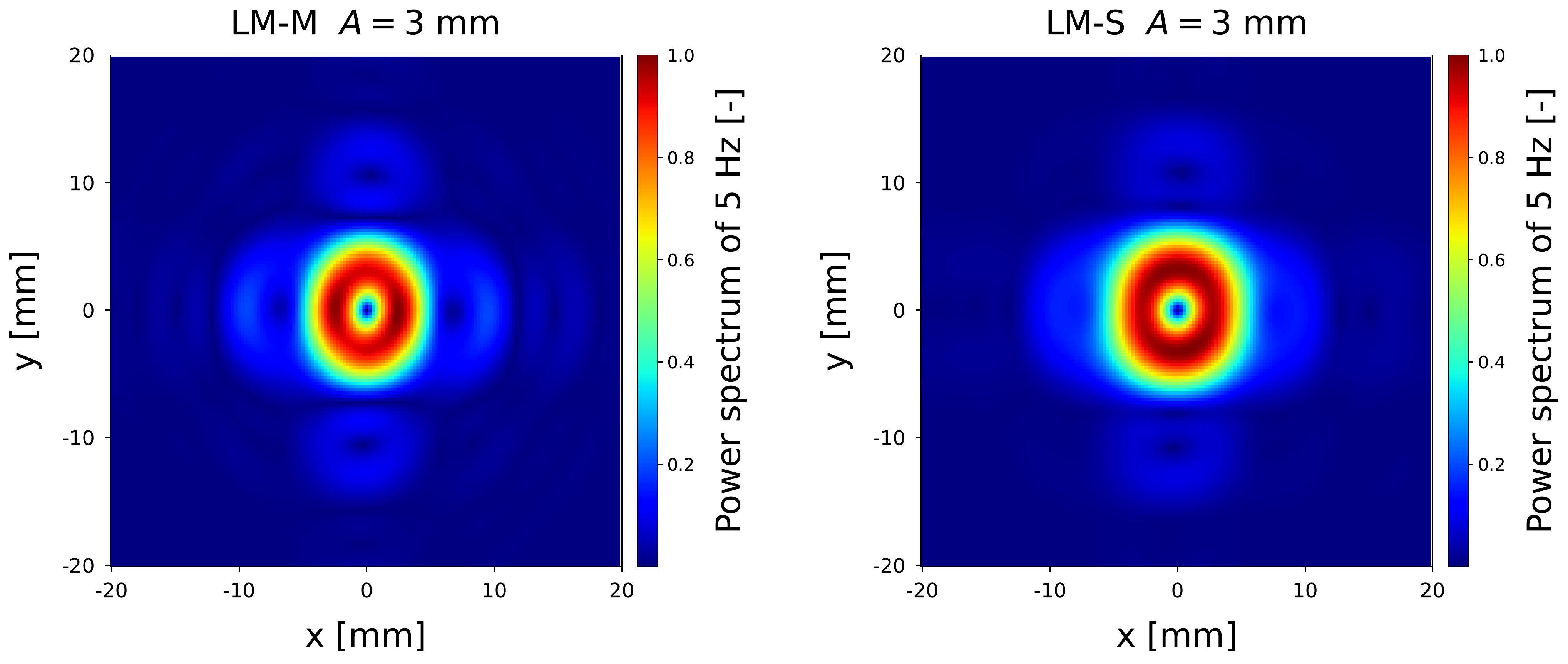}%
    \caption{Simulated 5 Hz-power spectrum distribution of time variation of radiation pressure produced by LM-M and LM-S at 5 Hz. The power spectrum distribution was obtained by simulating the time variation of the radiation pressure at each point in the stimulus area and Fourier transforming the time variation. These values were normalized.}
    \label{fig:Fig/Discussion/PowerSpectrumDistribuWith3mm.pdf}
\end{figure}

\subsection{Static Pressure Sensation at Finger Pad\label{sec: Discussion: Static Pressure Sensation}}
The results of Experiment 1 showed that LM-M and -S at 5 Hz can produce a non-vibratory pressure sensation on a finger pad. Moreover, with stimulus radii of $\StimulusRadius = 2, 3$~mm, the movement sensations were barely perceivable, and the pressure sensation was well static. The vibration sensation of the LM at 5 Hz was 4 or less in all conditions except LM-M with $\StimulusRadius = 5$~mm. For $\StimulusRadius = 2, 3$, the movement sensations of the LM-M were 2 or less. 

The results of Experiment 3 also showed that the perceived intensity of the pressure sensation on the finger pad was perceptually comparable to 0.16~N or more physical contact force on average. With the lowest vibration and movement sensation (LM-M with $\StimulusRadius = 3$~mm), the perceived force was 0.24~N, which was 10.9 times the radiation pressure at the focus presented in the setup.

However, in Experiment 3, extremely low and high forces were identified. For the LM-M with $\StimulusRadius = 3$~mm, the minimum and maximum values were 0 and 1.22~N, respectively. Note that the participant who answered 0~N could perceive the LM-M stimulus with $\StimulusRadius = 3$~mm. Since the answered equivalent force is less than 0.01~N, which is the minimum measurable force of the force gauge, the force is recorded as 0~N.
This large difference in perceived force could be attributed to the tactile receptor adaptation to the pushing stimulus presented by the force gauge. The pushing force is static, and the perceived intensity of such stimulus gradually weakens with stimulus duration owing to SA-I (slowly-adaptive type I) tactile receptor adaptation~\cite{nafe1941nature,iggo1969structure}. The previous study shows that the adaptation time for a static force of 0.175 N with a stimulus area of 50.18 mm$^2$ was 5.6 s~\cite{nafe1941nature}. As the stimulus duration of 2 s in Experiment 3 was long compared to the adaptation time, the perceived intensity of the pushing force may be greatly changed over 2~s. In such a situation, the answered force (shown in Fig.~\ref{fig:Fig/Ex3/PerceivedForce.pdf}) depends on the comparison timing (employment timing of the perceived intensity of the pushing force). For example, we considered that the comparison timing of the participants answering an extremely high force was late (nearly 2~s). When the timing is late, the perceived intensity of the pushing force is weak, resulting in a high answered force. Conversely, the comparison timing of the participants answering an extremely low force could be fast. We also considered that the individual difference in the adaptation speed could affect the answered equivalent force. When the adaptation speed is fast, the perceived intensity of the pushing force rapidly weakens over the period of 2~s, resulting in a high answered force. The evaluation of the individual differences in adaptation speed is important for future work.

The experimental results also indicated that the perceived intensity of the LM-M stimulus with $\StimulusRadius = 2$~mm was extremely weak. In Experiments 1 and 2, eight participants could not perceive the LM-M stimulus with $\StimulusRadius = 2$~mm. We considered that the weakness is because the circumference with a 2~mm radius and the length of the curved line-shaped stimulus distribution used in LM-M (9~mm) were almost the same. As an exception, in Experiment 3, only one participant could not perceive the LM-M stimulus with $\StimulusRadius = 2$~mm, and the average perceived force was 0.16~N. This difference could be attributed to the difference in the presentation time of the LM stimulus~\cite{vallbo1984properties}. In Experiment 3, the stimulus duration was 2~s, but in Experiments 1 and 2, the participants continued to be presented with the LM stimulus without any time limit. Therefore, in most participants in Experiments 1 and 2, their SA-I tactile receptors completely adapted to the LM-M stimulus, and they could not perceive the stimulus. 

\subsection{Perceived Curvature}
Since eight participants could not perceive the LM-M with $\StimulusRadius = 2$~mm in Experiments 1 and 2, we excluded it from the following discussions.

Experiments 1 and 2 suggest that a circular LM with $\StimulusRadius = 2$--$4$~mm can render a contact sensation with a convex surface with radii of 2--4~mm. As described in Section~\ref{sec: Discussion: Static Pressure Sensation}, particularly for the $\StimulusRadius = 2, 3$~mm, the contact sensation was well static. In the LM at 5 Hz with $\StimulusRadius = 2$--$4$~mm, the convex score was significantly higher than the concave score ($\pValue < 0.05$). In LM-M with $\StimulusRadius = 4$~mm and LM-S with $\StimulusRadius = 2$~mm, the convex score was significantly higher than the flat score ($\pValue < 0.05$). The perceived radii for LM-M with $\StimulusRadius = 3$~mm and LM-S with $\StimulusRadius = 4$~mm were 2 and 4, respectively. The participants' comments suggest that a convex sensation was rendered. Four participants commented that they sometimes felt in contact with sharp or rounded objects.

However, in some cases, participants found it difficult to determine whether the perceived contact shape was convex or flat. Two of the participants commented that this determination was difficult. Moreover, no significant differences were observed between the convex scores for the LM-S at 5 Hz with $\StimulusRadius = 3$~mm and LM-M with $\StimulusRadius = 3$~mm. The difficulty of the discrimination would not be attributed to the curvature of the perceived convex surface since the threshold of the curvature radius of the flat-convex discrimination is as large as 20.4 cm~\cite{goodwin1991tactile}. The maximum curvature radius of the convex surface image used in the experiment is 6 mm and is lower than the threshold. In the future, we will explore the reason for the difficulty of the discrimination. We will also quantitatively evaluate the curvature of the perceived surface and explore a control method for the curvature.

The curvature of the convex and concave images varied with respect to the LM radius $\StimulusRadius$. We conjectured that the effect of the variation on the curvature evaluation was minor since the variation range was small (2--6 mm). Even when $\StimulusRadius$ was the largest (6 mm), it was still smaller than the threshold of the curvature radius (20.4 cm) in which humans can discriminate against a flat and convex surface~\cite{goodwin1991tactile}.

In the LM at 5 Hz with $\StimulusRadius = 2$--$4$~mm, all concave scores were less than 2, and a concave sensation was not perceived. We considered that the periphery of the LM was hardly perceived in the radius range as three participants commented that they had high concave scores when they strongly perceived the perimeter of the stimulus. The time-averaged radiation pressure distribution of the LM was consistent with this consideration. Fig.~\ref{fig:Fig/Discussion/RadiationPressureDistribu.pdf} shows the simulated time-averaged pressure distribution. The simulation setup is the same as that shown in Fig.~\ref{fig:Fig/Equipment/SetupAUTD.pdf}. The results indicate that the periphery of the LM is the peak of the time-averaged radiation pressure only above a radius of 5 mm, where the concave score is high.

We finally confirmed that the perceived curvature did not match the 5 Hz-vibration intensity distribution produced by the LM at 5 Hz. Fig.~\ref{fig:Fig/Discussion/PowerSpectrumDistribuWith3mm.pdf} shows the simulated distribution of the 5 Hz vibration intensity (power spectrum) produced by 5 Hz LM-M and -S with $\StimulusRadius = 3$ mm. The power spectrum distribution was obtained by simulating the time variation of the radiation pressure at each point in the stimulus area and Fourier transforming the variation. The simulation setup is the same as that shown in Fig.~\ref{fig:Fig/Equipment/SetupAUTD.pdf}. The results showed that the physical intensity of the 5 Hz vibration was the highest on the focal orbits and did not match the perceived curvature. With $\StimulusRadius = 3$~mm, the 5 Hz LM-M and -S were perceived as contact with a convex surface. However, even under these conditions, the peaks of vibration intensity formed a circle, which is a contact shape with concave. In the future, we will investigate the relationship between perceived curvature and vibration intensity distribution by measuring or simulating the skin displacement as in previous studies~\cite{chilles2019laser,frier2022simulating}.

\subsection{Comparison of LM-M and LM-S at 5 Hz}
The results of Experiment 1 showed that the movement sensation of LM is suppressed by using multi foci (LM-M). With $\StimulusRadius = 3, 4, 5$, the movement sensation of the LM-M was significantly lower than that of the LM-S at 5 Hz. We considered that the suppression was because the simultaneously stimulated area of LM-M was wider than that of LM-S.

The LM-M was perceived to be smaller than the LM-S at 5 Hz. For $\StimulusRadius = 3, 4$~mm, the perceived size of LM-M was significantly smaller than that of LM-S at 5 Hz ($\pValue < 0.05$). This trend is consistent with the difference in the size of the time-averaged radiation pressure distribution. The simulation results (Fig.~\ref{fig:Fig/Discussion/RadiationPressureDistribu.pdf}) indicate that the time-averaged distributions of LM-M with $\StimulusRadius = 3, 4$~mm were smaller than those of LM-S. 

There were no huge differences between LM-M and LM-S at 5 Hz in the perceived intensity. Except for $\StimulusRadius = 2$~mm, there were no significant differences ($\pValue > 0.05$). However, the literature~\cite{shen2023multi} reported that multi-point STM were perceived as weaker than single-point STM. The difference could be attributed to the large focus spacing in the STM, causing the drop of the radiation force at each focus. The focus spacing of the STM was over 6 cm. In such a situation, where each focus is completely separated, the radiation force at each focus must be significantly dropped. The focus spacing of LM-M ($\MultiSpace$) was 3 mm, and the foci were not separated (Fig.~\ref{fig:Fig/Equipment/SimulatedRadiationPressure_simple.pdf}). The un-separated focus could result in the same perceived intensity between LM-M and LM-S.

\subsection{Comparison of Movement Sense Between LM Frequencies}
Experiment 1 showed the movement sensation of the LM at 25 Hz was significantly lower than that of the LM at 5 Hz for $\StimulusRadius = 3, 4, 5, 6$~mm. This could be because the focus speed at $\LmFreq = 25$ Hz was too fast to perceive the movement differently from the vibration. It is consistent with the previous study~\cite{freeman2021perception}. They found that the focal movement was not perceivable when the frequency of the circular STM with a diameter of 4--7~cm was over 18 Hz.

The results also showed that rendering a convex was difficult with the vibratory sensation produced by ultrasound. As the vibration score of 25 Hz LM-S was 6 or higher, this stimulus evoked a vibratory sensation. In the LM at 25 Hz, the flat score is the highest for all radii and was significantly greater than the convex score. One participant commented that the contact shape often felt flat when vibration was perceived.

\subsection{Limitation}
This study has three limitations. First, the synchronization (time delay and spatial alignment) between the visual marker and the tactile stimulus was not fully guaranteed. However, the participants' comments suggest this synchronization was not a major problem. All participants reported that they recognized the visual marker location and perceived a tactile stimulus when touching the marker with no delay.

Second, the actual shape and movement of the focus on the finger are not measured. In the future, we will measure the spatiotemporal pattern of the focus on the finger pad using a thermocamera~\cite{onishi2022two} and compare it with the perceived shape.

Third, although the focus position was continuously moved in the LM, the position was fixed in the radiation force measurement (in Section~\ref{sec: Measurement of Radiation Force}). Since the radiation force would be dropped when the focus was moved~\cite{suzuki2020reducing}, the amplification factor 10.9 times we evaluated, a ratio of the radiation force to the perceived intensity, may be underestimated.

\section{Conclusion}
In this study, we verified that ultrasound radiation pressure distribution, which spatiotemporally varies at 5 Hz, can provide a static pressure sensation on a finger pad. We also demonstrated that the pressure sensation on the finger pad was perceived as a static contact sensation with a convex surface. In the experiment, four ultrasound focal points were presented on the finger pads of the participant and they were simultaneously rotated in a circle at 5 Hz. When the radius of the focal trajectory was 3~mm, the perceived vibration and movement sensations were the lowest, 1.5 and 2 out of 7 on average, respectively. The perceived intensity of this evoked pressure sensation was equivalent to a 0.24 N physically constant force lasting for 2~s, which is 10.9 times the physically presented radiation force at the focus. Under the most static condition, the pressure sensation was perceived as a contact sensation on a convex surface with a radius of 2~mm. The average perceptual similarity was 5 out of 7. 

From these results, we conclude that focused ultrasound can render a static contact sensation at a finger pad with a small convex surface. This contact sensation rendering enables the noncontact tactile reproduction of a static-fine uneven surface. In the future, we will investigate curvature control of the rendered convex surface.

Finally, all the data used in this paper are available at Open Science Framework (URL: https://osf.io/e7x2b/).

% or


\ifCLASSOPTIONcaptionsoff
  \newpage
\fi



\bibliographystyle{IEEEtran}
\bibliography{ref}

\begin{IEEEbiography}[{\includegraphics[width=1in,height=1.25in,clip,keepaspectratio]{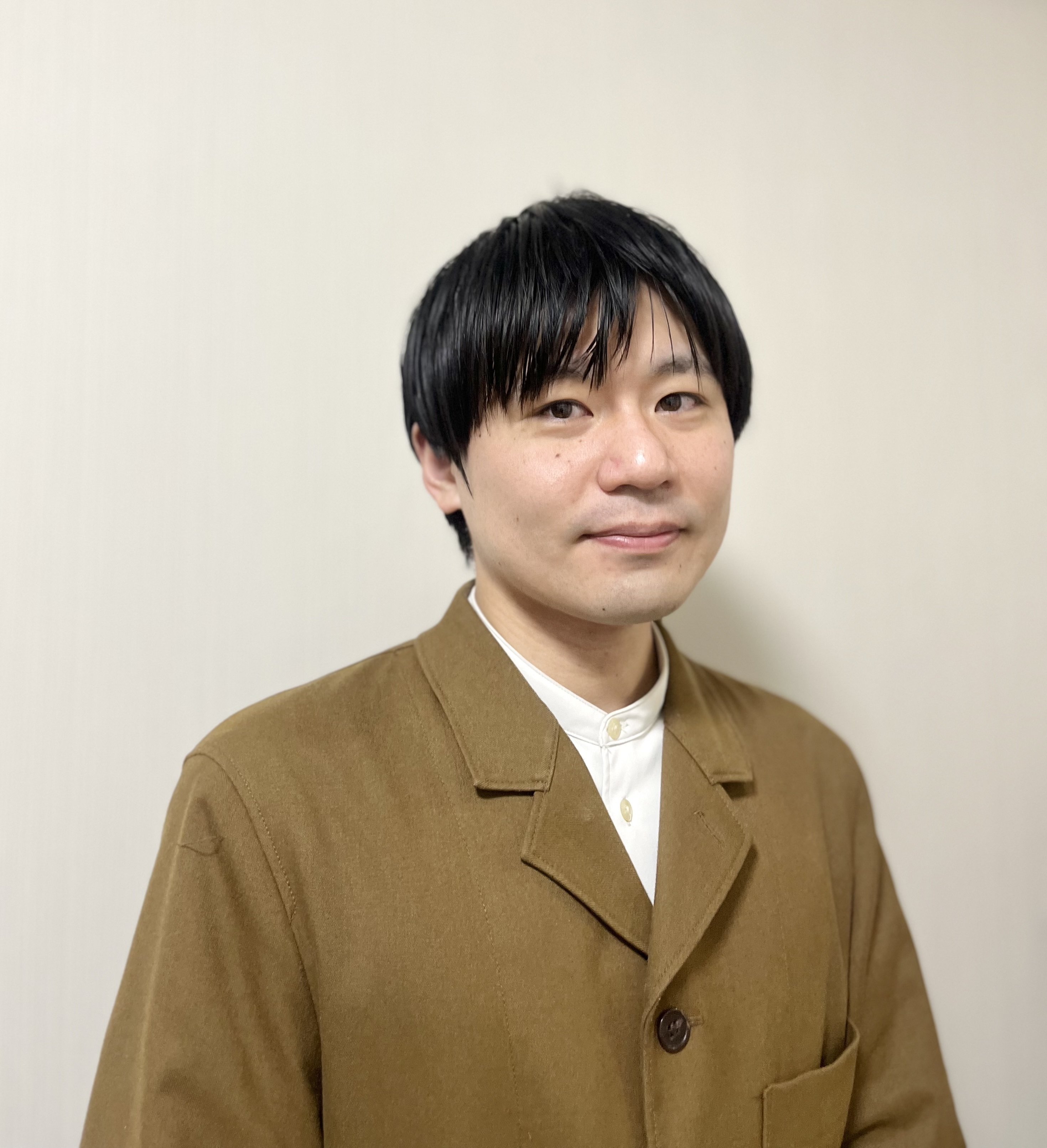}}]{Tao Morisaki}
Tao Morisaki is a researcher at NTT Communication Science Laboratories. His previous affiliation was The University of Tokyo, Japan. He received the M.S. degree in 2020 and the Ph.D. degree in 2023 from the Department of Complexity Science and Engineering, The University of Tokyo. His research interests include haptics, ultrasound midair haptics, and human-computer interaction. He is a member of IEEE and VRSJ.
\end{IEEEbiography}
\begin{IEEEbiography}[{\includegraphics[width=1in,height=1.25in,clip,keepaspectratio]{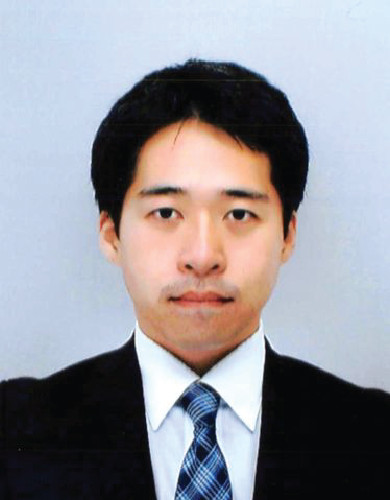}}]
{Masahiro Fujiwara}
He is a lecturer at the Faculty of Science and Technology, Nanzan University, Japan. His previous affiliation was the University of Tokyo, Japan. He received the PhD degree in Information Science and Technology from the University of Tokyo in 2015. His research interests include information physics, haptics, non-contact sensing and those application systems. He is a member of the Society of Instrument and Control Engineers.
\end{IEEEbiography}
\begin{IEEEbiography}[{\includegraphics[width=1in,height=1.25in,clip,keepaspectratio]{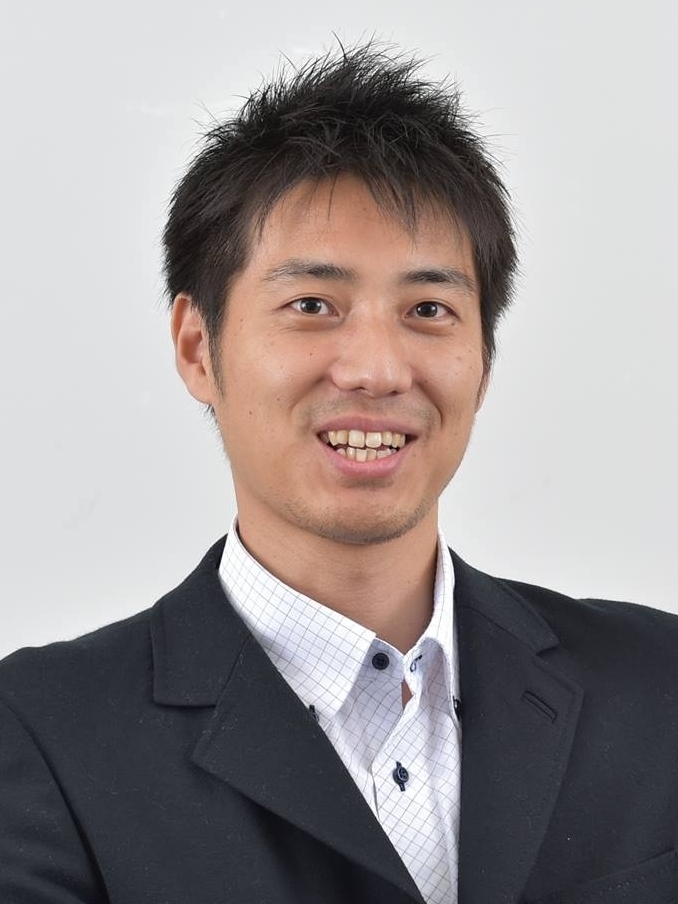}}]{Yasutoshi Makino}
Yasutoshi Makino is an associate professor in the Department of Complexity Science and Engineering in the University of Tokyo. He received his PhD in Information Science and Technology from the Univ. of Tokyo in 2007. He worked as a researcher for two years in the Univ. of Tokyo and an assistant professor in Keio University from 2009 to 2013. From 2013 he moved to the Univ. of Tokyo as a lecture, and he is an associate professor from 2017. His research interest includes haptic interactive systems.
\end{IEEEbiography}
\begin{IEEEbiography}[{\includegraphics[width=1in,height=1.25in,clip,keepaspectratio]{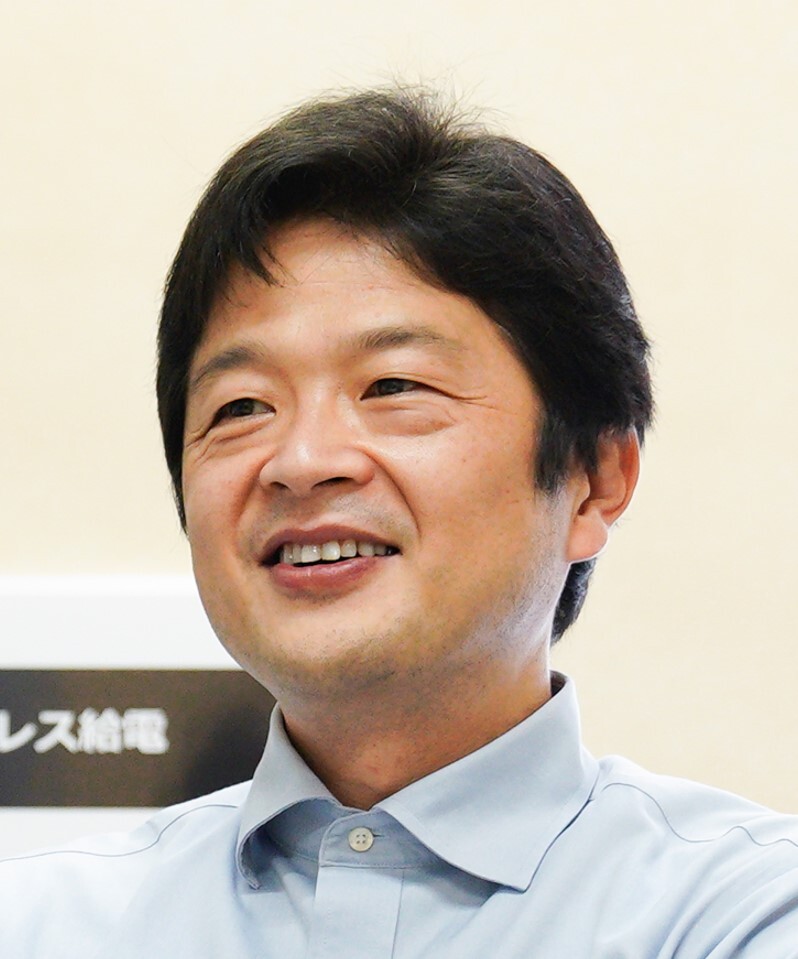}}]{Hiroyuki Shinoda}
Hiroyuki Shinoda is a Professor at the Graduate School of Frontier Sciences, the University of Tokyo. After receiving a Ph.D. in engineering from the University of Tokyo, he was an Associate Professor at Tokyo University of Agriculture and Technology from 1995 to 1999. He was a Visiting Scholar at UC Berkeley in 1999 and was an Associate Professor at the University of Tokyo from 2000 to 2012. His research interests include information physics, haptics, mid-air haptics, two-dimensional communication, and their application systems. He is a member of SICE, IEEJ, RSJ, JSME, VRSJ, IEEE and ACM.
\end{IEEEbiography}
\vfill

%\vfill


\end{document}